\documentclass[%
 reprint,
 superscriptaddress,
 groupedaddress,
 amsmath,
 amssymb,
 aps,
 pre,
]{revtex4-2}

\usepackage{dcolumn}  
\usepackage{hyperref}  
\usepackage{bm}  
\usepackage{amsthm}
\usepackage{newtxtext}
\usepackage{mathtools}  
\usepackage{cases}
\usepackage{centernot}
\usepackage{blindtext}
\usepackage{enumerate}

\usepackage{kotex}
\usepackage{lipsum}
\usepackage{booktabs}

\begin{document}

\preprint{APS/123-QED}

\title{Hyperdiffusion of Poissonian run-and-tumble particles in two dimensions}

\author{Yurim Jung}%
 \email{dmavy42@postech.ac.kr}
 
\affiliation{%
Department of Physics, Pohang University of Science and Technology, Pohang 790-784, Korea
}%




\date{\today}

\begin{abstract}
We study non-interacting Poissonian run-and-tumble particles (RTPs) in two dimensions whose velocity orientations are controlled by an arbitrary circular distribution $Q(\phi)$.
RTP-type active transport has been reported to undergo localization inside crowded and disordered environments, yet its non-equilibrium dynamics, especially at intermediate times, has not been elucidated analytically.
Here, starting from the standard (one-state) RTPs, we formulate the localized (two-state) RTPs by concatenating an overdamped Brownian motion in a Markovian manner.
Using the space-time coupling technique in continuous-time random walk theory, we generalize the Montroll-Weiss formula in a decomposable form over the Fourier coefficient $Q_{\nu}$ and reveal that the displacement moment $\left \langle \mathbf{r}^{2\mu}(t) \right \rangle$ depends on finite angular moments $Q_{\nu}$ for $|\nu|\leq \mu$.
Based on this finding, we provide (i) the angular distribution of velocity reorientation for one-state RTPs and (ii) $\left \langle \mathbf{r}^{2}(t) \right \rangle$ over all timescales for two-state RTPs.
In particular, we find the intricate time evolution of $\left \langle \mathbf{r}^{2}(t) \right \rangle$ that depends on initial dynamic states and, remarkably, detect hyperdiffusive scaling $\left \langle \mathbf{r}^{2}(t) \right \rangle \propto t^{\beta(t)}$ with an anomalous exponent $2<\beta(t)\leq 3$ in the short- and intermediate-time regimes.
Our work suggests that the localization emerging within complex systems can increase the dispersion rate of active transport even beyond the ballistic limit.
\end{abstract}


\maketitle

\section{\label{sec:introduction}Introduction}

Active particles, known as self-propelled particles (SPPs), have been extensively studied owing to their characteristics beyond thermodynamic equilibrium \cite{marchetti2013hydrodynamics, bechinger2016active}, such as collective behaviors of swarming \cite{vicsek1995novel}, self-organization \cite{bressloff2013stochastic}, pattern formation \cite{bar2020self}, and phase separation \cite{fily2012athermal, cates2015motility}.
Furthermore, even at a non-interacting single particle level, they perform a directed motion powered by extra energy sources and manifest superdiffusive mean-squared displacement (MSD) $\left \langle \mathbf{r}^{2}(t) \right \rangle \propto t^{\beta(t)}$ with an anomalous exponent $\beta(t) > 1$.
However, these ballistic agents may undergo intermittent \textit{localization} inside crowded and disordered environments and reach even more peculiar dynamics where multiple motion states coexist \cite{bressloff2013stochastic}.
For instance, motor proteins and/or cargo complexes exhibit jiggling or cycling motions at the junctions of biopolymer networks, e.g., microtubule \cite{balint2013correlative, chen2015memoryless}, actin \cite{scholz2016cycling}, and mitochondria meshes \cite{chan2018microtubule} in cells.
Our aim is to analytically describe such \textit{localized} SPPs in a two-dimensional ($2$D) space, starting from the canonical diffusion model of run-and-tumble particles (RTPs).

RTPs are originally developed to formulate the motility patterns of microorganisms \cite{berg1972chemotaxis}, which display ballistic motions (run) that sporadically alter the direction (tumble) by a turning angle $\phi$.
Notably, a mean absolute turning angle $\left \langle |\phi| \right \rangle$ may differ depending on the bacteria's swimming strategy, for example, $\left \langle |\phi| \right \rangle\approx68^{\circ}$ for \textit{E. coli} \cite{berg1993random}, $180^{\circ}$ for \textit{P. haloplanktis} \cite{johansen2002variability}, switching between $90^{\circ}$ and $180^{\circ}$ for \textit{V. alginolyticus} \cite{xie2011bacterial}. 
To meet the observations, the model establishes a circular distribution $Q(\phi)$ and calculates the angle-dependent quantities, such as the MSD \cite{lovely1975statistical}, velocity autocorrelation function \cite{taktikos2013motility}, and (excess) kurtosis \cite{villa2020run, sevilla2020two}.
On the other hand, a probability density function (PDF) of displacement $P(\mathbf{r},t)$ has not been clearly resolved in a continuum except for the most trivial case with $Q(\phi)=1/(2\pi)$ \cite{martens2012probability, santra2020run}.
%
Besides, even though numerous multi-state RTPs \cite{bressloff2011quasi, thiel2012anomalous, hafner2016run, malakar2018steady, perez2019bacteria, shaebani2022kinematics} have been proposed, an elementary two-state model, interrupted by diffusive motion in the $2$D plane, remains veiled in terms of MSD at all timescales.

An alternative theoretical model for SPPs is active Brownian particles (ABPs), where the propagation direction of the tracer undergoes rotational diffusion encoded in Langevin dynamics.
The MSD of overdamped ABPs in two dimensions reads \cite{basu2018active}
\begin{equation}
    \left \langle \mathbf{r}^{2}(t) \right \rangle = \frac{2v^{2}}{D_{\mathrm{R}}}t+\frac{2v^{2}}{D_{\mathrm{R}}^{2}}(e^{-D_{\mathrm{R}}t}-1),
    \label{eq:abp-msd}
\end{equation}
where $D_{\mathrm{R}}$ denotes the rotational diffusivity and $v$ is the speed of the tracer.
Interestingly, the identical form of MSD to Eq.~\eqref{eq:abp-msd} is reproducible for RTPs \cite{comments-msd}; accordingly, numerous comparative studies on ABPs and RTPs have been conducted \cite{solon2015active, kurzthaler2018probing, dauchot2019dynamics}.
For instance, Ref.~\cite{solon2015active} have been revealed that two dynamics are distinguishable when each system is heavily confined to an external harmonic trap.

To formulate the RTPs with localization, we follow transport equations based on the continuous-time random walk (CTRW) \cite{montroll1965random} rather than applying the Fokker-Planck equations \cite{malakar2018steady}.
In particular, we benchmark the L\'evy walk \cite{klafter2011first, zaburdaev2015levy}, an active process where a linear coupling $l=v\tau$ between the tracer's step length $l$ and its interarrival time $\tau$ is assumed via a constant speed $v$.
The model adopts L\'evy statistics and renders out-of-Brownian properties, such as breakdown of the central limit theorem (CLT) and weak ergodicity breaking \cite{zaburdaev2015levy}.
Nonetheless, the L\'evy-type motion is prevalent in superdiffusive phenomena, e.g., photon scattering in L\'evy glass \cite{barthelemy2008levy, burresi2012weak}, motility pattern of T cells \cite{harris2012generalized}, and albatross flight \cite{edwards2007revisiting}.
This study focuses on the mathematical flexibility of CTRW \cite{klafter1994levy, froemberg2015asymptotic, zaburdaev2013space, zaburdaev2016superdiffusive} and exploits its space-time coupling technique instead of adopting the entire heavy-tailed process.

Let us specify the scope and purpose of this paper.
We study non-interacting Poissonian RTPs \cite{detcheverry2017generalized}, whose orientation of self-propulsion is manipulated by $Q(\phi)$ on the $2$D plane; here, we consider two cases: particles without localization (one-state RTPs) and with localization (two-state RTPs).
Our aim is to provide a physical understanding of the localized SPPs via our two-state model, where the spatially trapped state displays a passive Brownian motion with diffusivity $D$.
Moreover, for one-state RTPs, we extend the previous analysis \cite{lovely1975statistical, taktikos2013motility, villa2020run, sevilla2020two} to measurable quantities, e.g., an angular distribution of velocity reorientation $\mathcal{Q}(\Theta , t)$, whereby we discover highly dissimilar angle relaxations between ABPs and RTPs.
For two-state RTPs, we derive the MSD $\left \langle \mathbf{r}^{2}(t) \right \rangle$ and detect the intricate time evolution of $\beta(t)$, which strongly depends on initial states and thus implies a connection to heterogeneous (site-dependent) diffusion processes (HDPs) \cite{cherstvy2013population, grebenkov2018heterogeneous, xu2020heterogeneous}.
Remarkably, for a sufficiently small $D/v^2$, we report a transient \textit{hyperdiffusion} with $2<\beta(t)\leq 3$ as a macroscopic phenomenon that arises from the particles gradually escaping the localized state.

The paper is structured as follows.
In Sec.~\ref{sec:formalism}, we elaborate our theoretical models utilizing the space-time coupled CTRW.
Then, in Secs.~\ref{sec:results-one} and \ref{sec:results-two}, we investigate the averaged quantities of one- and two-state RTPs in terms of the PDF, MSD, non-Gaussianity, and time-dependent cosine moment of $\mathcal{Q}(\Theta , t)$.
In Sec.~\ref{sec:discussion}, we discuss the hyperdiffusion phenomenon in detail, and last, we summarize the critical results of this study in Sec.~\ref{sec:conclusion}.

\section{\label{sec:formalism}Formalism}

In this section, we reformulate the RTP dynamics in the $2$D plane under the space-time coupled CTRW framework \cite{klafter2011first}.
Consider a particle performing a ballistic motion with a constant velocity $\mathbf{v}=v\mathbf{\hat{n}}$ with orientation $\mathbf{\hat{n}}=\cos\theta\mathbf{\hat{x}}+\sin\theta\mathbf{\hat{y}}$ for a duration $t_{1}$ drawn from an exponential distribution $\psi_{a}(t)=\gamma_{a}\exp(-\gamma_{a}t)$.
The particle halts after the completion of run state and instantly assumes a new velocity orientation $\theta$ sampled from a uniform distribution $U(\theta)=1/(2\pi)$.
These consecutive iterations between run and reorientation, known as tumble \cite{taktikos2013motility}, yield the following governing equations:
\begin{align}
    \zeta (\mathbf{r}, t) = \int_{0}^{t} dt_{1} \int_{\mathbb{R}^{2}} \zeta(\mathbf{r}-\mathbf{r}_{1}, t-t_{1}) \psi_{a}(t_{1}) p_{a}(\mathbf{r}_{1}|t_{1}) d^{2}\mathbf{r}_{1} \nonumber
    \\ + \delta(\mathbf{r})\delta(t), \nonumber
    \\
    P(\mathbf{r}, t) = \int_{0}^{t} dt_{1} \int_{\mathbb{R}^{2}} \zeta(\mathbf{r}-\mathbf{r}_{1}, t-t_{1}) \Psi_{a}(t_{1}) p_{a}(\mathbf{r}_{1}|t_{1}) d^{2}\mathbf{r}_{1}, 
    \label{eq:rtp-convention-prt}
\end{align}
where a conditional PDF $p_{a}(\mathbf{r}_{1}|t_{1})=\delta (|\mathbf{r}_{1}|-vt_{1})/(2\pi vt_{1})$ encodes a space-time coupling to travel a distance $vt_{1}$ for a given $t_{1}$.
Here, $\zeta (\mathbf{r}, t)$ represents the PDF of arriving at position $\mathbf{r}$ at time $t$ while completing the run state and $\delta(\mathbf{r})\delta(t)$ is set as the initial condition.
For the last step, we define a survival probability \cite{klafter2011first} $\Psi_{a}(t_{1}):=\int_{t_{1}}^{\infty}\psi_{a}(\tau)d\tau$ to add particles incomplete in the run state but have reached $\mathbf{r}$ at $t$.
We note that Eq.~\eqref{eq:rtp-convention-prt}, i.e., the Van Hove function of RTPs with $U(\theta)$ (or conventional RTPs), is solvable in the Fourier-Laplace space (see Appendix~\ref{app:sec:functions-one}).

The MSD of the conventional RTPs is given as \cite{villa2020run, santra2020run}  
\begin{equation}
    \left \langle \mathbf{r}^{2}(t) \right \rangle = 
    \frac{2v^{2}}{\gamma_{a}}t+\frac{2v^{2}}{\gamma_{a}^{2}}(e^{-\gamma_{a}t}-1),
    \label{eq:rtp-convention-msd}
\end{equation}
where the rate parameter $\gamma_{a}$ replaces the rotational diffusion constant $D_{\mathrm{R}}$ of the MSD of the ABPs written in Eq.~\eqref{eq:abp-msd} \cite{cates2013active}.

\subsection{\label{sec:formalism-one}One-state RTPs}

\begin{figure}[!t]
\centering
\includegraphics[keepaspectratio=true, width=8.6cm, height=150cm]{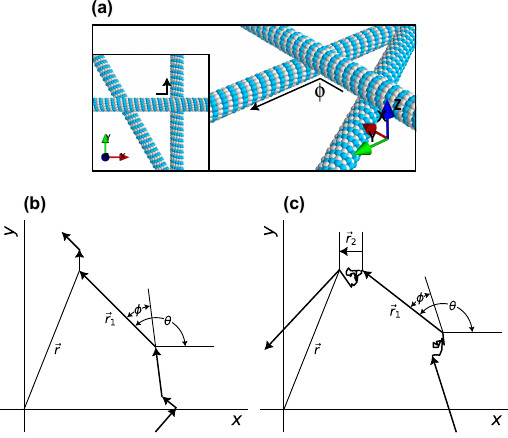}
\caption{
Trajectories of RTPs (bold lines) in a quasi-$2$D geometry.
We consider two cases: RTPs (b) without localization and (c) with localization.
(a) The schematics of a disordered system, e.g., biofilament networks in live cells.
The active tracer may experience a local diffusive motion in intersections.
An active-to-active turning angle $\phi$ is drawn from $Q(\phi)$.
(b) One-state RTPs.
The tracer at $\mathbf{r}-\mathbf{r}_{1}$ at time $t-t_{1}$ rotates by $\phi$ and performs ballistic motion with $\mathbf{r}_{1}=vt_{1}\mathbf{\hat{n}}$ to reach $\mathbf{r}$ at time $t$.
We denote $\mathbf{\hat{n}}=\cos\theta\mathbf{\hat{x}}+\sin\theta\mathbf{\hat{y}}$.
(c) Two-state RTPs.
The tracer at $\mathbf{r}-\mathbf{r}_{1}-\mathbf{r}_{2}$ at time $t-t_{1}-t_{2}$ exhibits the identical directed motion to (b) by memorizing the previous active orientation $\theta-\phi$.
After reaching $\mathbf{r}-\mathbf{r}_{2}$ at time $t-t_{2}$, it immediately undergoes localization, i.e., overdamped Brownian motion, yielding displacement $\mathbf{r}_{2}$ for duration $t_{2}$.
The tracer arrives at $\mathbf{r}$ at time $t$.
}
\label{fig:formalism}
\end{figure}

We extend Eq.~\eqref{eq:rtp-convention-prt} for the general case where the next velocity orientation is not independent of the previous.
In detail, when a particle reaches $\mathbf{r}-\mathbf{r}_{1}$ while headed in the direction $\theta-\phi$, we force it to perform a linear motion in the $\theta$ direction for duration $t_{1}$ by introducing a turning angle $\phi$ from a circular distribution $Q(\phi)$ [see Fig.~\ref{fig:formalism}(b)].
Accordingly, the particle precisely arrives at position $\mathbf{r}$ at time $t$.
Formally, we write
\begin{align}
    \zeta (\mathbf{r}, \theta, t) 
    =\int_{0}^{t} dt_{1} \int_{-\pi}^{\pi} d\phi \int_{\mathbb{R}^{2}} \zeta(\mathbf{r}-\mathbf{r}_{1}, \theta - \phi, t-t_{1}) \nonumber
    \\
    \times Q(\phi) \psi_{a}(t_{1}) p_{a}(\mathbf{r}_{1}|\theta,t_{1}) d^{2}\mathbf{r}_{1} + \delta(\mathbf{r})\delta(t)P_{0}(\theta),
    \label{eq:one-zeta}
\end{align}
where the PDF $p_{a}(\mathbf{r}_{1}|\theta,t_{1})=\delta (|\mathbf{r}_{1}|-vt_{1})\delta (\theta_{1}-\theta)/(vt_{1})$.
Here, we denote $\zeta (\mathbf{r}, \theta, t)$ as the particle density of reaching $\mathbf{r}$ with the velocity orientation $\theta$ at time $t$ just completing the ballistic phase and set $\delta(\mathbf{r})\delta(t)P_{0}(\theta)$ as the initial condition.
Note that two random variables $\phi$ and $t_{1}$ are independent, and we restrict an initial velocity orientation distribution to $P_{0}(\theta)=1/(2\pi)$.
Consequently, we obtain the propagator
\begin{align}
    P(\mathbf{r}, t)
    =\int_{-\pi}^{\pi} d\theta \int_{0}^{t} &dt_{1}\int_{-\pi}^{\pi}d\phi \int_{\mathbb{R}^{2}} \zeta(\mathbf{r}-\mathbf{r}_{1}, \theta - \phi, t-t_{1}) \nonumber
    \\
    &\times Q(\phi) \Psi_{a}(t_{1}) p_{a}(\mathbf{r}_{1}|\theta,t_{1}) d^{2}\mathbf{r}_{1},
    \label{eq:one-prt}
\end{align}
where we marginalize the orientation variable $\theta$ to consider all particles residing at the final position $\mathbf{r}$ at time $t$.

Let us perform the Fourier transform ($\mathbf{r}\rightarrow \mathbf{k}$) and subsequently Laplace transform ($t\rightarrow s$) of Eq.~\eqref{eq:one-zeta}, which yield
\begin{align}
    \hat{\zeta} (\mathbf{k}, \theta, s) =
    &\sum_{n=-\infty}^{\infty} (i)^{-n}e^{in\psi}
    \mathcal{L} \{ \psi_{a}(t_{1})J_{n}(|\mathbf{k}|vt_{1}) \} e^{-in\theta} \nonumber \\
    &\hspace{0.5cm}
    \times \int_{-\pi}^{\pi} d\phi \hat{\zeta} (\mathbf{k}, \theta-\phi,s)Q(\phi) + P_{0}(\theta),
    \label{eq:one-zeta-jacobi-anger}
\end{align}
where $J_{n}(x)$ is the Bessel function of the first kind and $\mathcal{L} \{ \cdots \}$ abbreviates the Laplace transform (see Appendix~\ref{app:sec:transforms}).
Here, we decompose the dependence on orientation $\theta$ of $p_{a}(\mathbf{r}_{1}|\theta,t_{1})$ through the Jacobi-Anger expansion such that $\exp(-i\mathbf{k}\cdot\mathbf{v}t_{1}) = \sum_{n=-\infty}^{\infty} (-i)^{n} J_{n}(|\mathbf{k}|vt_{1}) e^{-in\theta} e^{in\psi}$ where $\psi:=\arctan(k_{y}/k_{x})$.
Furthermore, because the integral in Eq.~\eqref{eq:one-zeta-jacobi-anger} indicates a circular convolution [see Eq.~\eqref{app:eq:circular-convolution-definition}], we can treat the equation separable in ($\mathbf{k}, s$) and $\theta$.
Hence, inserting Fourier series expansions $\hat{\zeta} (\mathbf{k}, \theta,s)=\sum_{n=-\infty}^{\infty} \hat{\zeta}_{n}(\mathbf{k}, s)e^{in\theta}$ and $Q(\phi)=1/(2\pi) \sum_{n=-\infty}^{\infty} Q_{n}e^{in\phi}$, we simplify Eq.~\eqref{eq:one-zeta-jacobi-anger} as
\begin{equation}
    \hat{\zeta} (\mathbf{k}, \theta, s)
    =\sum_{m,n=-\infty}^{\infty} f_{n}(\mathbf{k},s) \hat{\zeta}_{m}(\mathbf{k},s) Q_{m} e^{i(m-n)\theta} + P_{0}(\theta), 
    \label{eq:one-zeta-hat}
\end{equation}
where
\begin{equation}
    f_{n}(\mathbf{k},s)=(i)^{-n}e^{in\psi}
    \mathcal{L} \{ \psi_{a}(t_{1})J_{n}(|\mathbf{k}|vt_{1}) \},
    \label{eq:one-fn}
\end{equation}
whose explicit expression is written in Eq.~\eqref{app:eq:one-fn}.

In the same manner, we calculate the Fourier-Laplace transform of $P(\mathbf{r}, t)$ in Eq.~\eqref{eq:one-prt}, which results in
\begin{equation}
    \hat{P}(\mathbf{k}, s) = \sum_{n=-\infty}^{\infty} 2\pi F_{n}(\mathbf{k},s)\hat{\zeta}_{n}(\mathbf{k},s) Q_{n}, 
    \label{eq:one-pks}
\end{equation}
where
\begin{equation}
    F_{n}(\mathbf{k},s)=(i)^{-n}e^{in\psi}
    \mathcal{L} \{ \Psi_{a}(t_{1})J_{n}(|\mathbf{k}| vt_{1}) \}.
    \label{eq:one-large-fn}
\end{equation}
We highlight that Eqs.~\eqref{eq:one-zeta-hat} and \eqref{eq:one-pks} generalize the Montroll-Weiss formula \cite{montroll1965random} in the decomposable form over Fourier coefficients $Q_{n}$ of an arbitrary circular distribution $Q(\phi)$.
Throughout the paper, we restrict ourselves to the exponential distribution $\psi_{a}(t)=\gamma_{a}\exp(-\gamma_{a}t)$.
With this assumption, we obtain an additional relation $F_{n}(\mathbf{k},s)=f_{n}(\mathbf{k},s)/\gamma_{a}$.

We solve Eq.~\eqref{eq:one-zeta-hat} on the Fourier basis, specifically by the following truncated matrix equation for a non-negative cutoff index $\mu$:
\begin{equation}
    (I-\mathcal{A}\mathcal{D})Z^{\mu} = E_{0},
    \label{eq:one-matrix-equation}
\end{equation}
where the column vector $Z^{\mu}=[Z_{-\mu}^{\mu }, \dots, Z_{0}^{\mu}, \dots, Z_{\mu}^{\mu}]^{\mathrm{T}}$ fulfills $\lim_{\mu\rightarrow \infty} Z_{\nu}^{\mu}(\mathbf{k},s)=\hat{\zeta}_{\nu}(\mathbf{k},s)$ for each Fourier mode index $\nu$ where $-\mu\leq \nu\leq \mu$.
Also, we introduce a Toeplitz matrix $\mathcal{A}=[\mathcal{A}_{mn}]$ with the component $\mathcal{A}_{mn}=f_{n-m}(\mathbf{k},s)$ for indices $m,n \in \left \{1, \dots, 2\mu+1 \right \}$ and a diagonal matrix $\mathcal{D}=\mathrm{diag}(Q_{-\mu}, \dots, Q_{0}, \dots, Q_{\mu})$.
Last, the initial condition $P_{0}(\theta)=1/(2\pi)$ gives the corresponding column vector $E_{0}=[0, \dots, 1/(2\pi), \dots, 0]^{\mathrm{T}}$.

For zero $\nu$, Eq.~\eqref{eq:one-matrix-equation} is rewritten as
\begin{equation}
    Z_{0}^{\mu} = \sum_{n=-\mu}^{\mu} 
    f_{n}(\mathbf{k},s) Z_{n}^{\mu}(\mathbf{k},s) Q_{n} + \frac{1}{2\pi},
    \label{eq:one-matrix-equation-zero}
\end{equation}
where the constant term on the right-hand side is derived from $P_{0}(\theta)=1/(2\pi)$.
In contrast, for nonzero $\nu$, we have
\begin{equation}
    Z_{\nu}^{\mu}=\sum_{n=-\mu}^{\mu} f_{n-\nu}(\mathbf{k},s) Z_{n}^{\mu}(\mathbf{k},s) Q_{n}.
    \label{eq:one-matrix-equation-nonzero}
\end{equation}
In the following section, we get $Z_{0}^{\mu}$ from Eq.~\eqref{eq:one-matrix-equation} by operating the inverse of a matrix $\mathcal{B}:=I-\mathcal{A}\mathcal{D}$ on the other side $E_{0}$.

With the help of $F_{n}(\mathbf{k},s)=f_{n}(\mathbf{k},s)/\gamma_{a}$, we can cut down Eq.~\eqref{eq:one-pks} as
\begin{equation}
    \hat{P}^{(\mu)} (\mathbf{k}, s) = 
    \frac{2\pi}{\gamma_{a}} \big(Z_{0}^{\mu} -\frac{1}{2\pi}\big)
    \label{eq:one-pks-mu}
\end{equation}
for a non-negative integer $\mu$.
In general, $\lim_{\mu \rightarrow \infty} \hat{P}^{(\mu)} (\mathbf{k}, s)=\hat{P} (\mathbf{k}, s)$ is satisfied.
Nonetheless, $\hat{P} (\mathbf{k}, s)$ can be equivalent to its approximation $\hat{P}^{(\mu)} (\mathbf{k}, s)$ if all Fourtier coefficients $Q_{\nu}$ vanish for $|\nu|>\mu$.
For instance, in the most trivial case of uniform tumbling, we yield $\mathcal{D}=\mathrm{diag}(0,\dots,1,\dots,0)$ from $Q(\phi)=1/(2\pi)$, and thus recover $\hat{P} (\mathbf{k}, s)$ of the conventional RTPs from $\hat{P}^{(0)} (\mathbf{k}, s)$ [see Eq.~\eqref{app:eq:convention-pks}].

Furthermore, without losing generality, we can calculate the $2\mu$th moment of displacement for any $Q(\phi)$ through
\begin{equation}
    \left \langle \mathbf{r}^{2\mu}(s) \right \rangle 
    = (-i)^{2\mu}\nabla_{\mathbf{k}}^{2\mu} \hat{P}^{(\mu)}(\mathbf{k},s)\big|_{\mathbf{k}=0},
    \label{eq:one-r2mu-del}
\end{equation}
whose Laplace inversion brings out our first main result,
\begin{equation}
    \left \langle \mathbf{r}^{2\mu}(t) \right \rangle 
    = \mathcal{F}_{\mu}(t;Q_{0}, Q_{\pm1}, \dots, Q_{\pm\mu}),
    \label{eq:one-r2mu-qmu-relation}
\end{equation}
which means that $\left \langle \mathbf{r}^{2\mu}(t) \right \rangle$ depends solely on Fourier coefficients $Q_{\nu}$ for $|\nu| \leq \mu$ and does not require the complete information of $Q(\phi)$.
Let us consider the polynomial expansion of $f_{n}$ and $Z_{n}^{\mu}$ at point $\mathbf{k}=0$.
We find that $|\mathbf{k}|^{|n|}$ is the lowest order term in $\mathbf{k}$ for both functions (here, we ignore their coefficients).
Hence, the lowest order term of $f_{n}Z_{n}^{\mu}$ in Eq.~\eqref{eq:one-matrix-equation-zero} becomes $|\mathbf{k}|^{|2n|}$.
Accordingly, the higher order terms for indices $|n|\geq\mu+1$ in Eq.~\eqref{eq:one-matrix-equation-zero} cannot contribute to $\left \langle \mathbf{r}^{2\mu}(s) \right \rangle $ (see also Appendix~\ref{app:sec:proof}).

Let us embody our RTP model with a circular distribution $Q(\phi)$ which satisfies reflective symmetry about $\phi=0$.
In this case, utilizing the $\mu$th cosine moment of $Q(\phi)$ denoted by
\begin{equation}
    \alpha_{\mu}:=\left \langle \cos \mu \phi \right \rangle,
    \label{eq:alpha-definition}
\end{equation}
we can shorten $\mathcal{D}=\mathrm{diag}(\alpha_{\mu},\dots,\alpha_{0},\dots,\alpha_{\mu})$ in Eq.~\eqref{eq:one-matrix-equation}.
More specifically, we adopt two circular distributions described by a concentration parameter $\kappa$: a cardioid (CD) and a wrapped Cauchy (WC) distribution \cite{mardia2000directional, ley2017modern} (see the definition in Appendix~\ref{app:sec:circular}).
For the CD, we report that the first cosine moment $\alpha_{1}=\kappa$ and $\alpha_{\mu\geq2}=0$ for $|\kappa|\leq1/2$.
For the WC, we have a relation $\alpha_{\mu}=\kappa^{\mu}$ for $0\leq \kappa <1$.
In Sec.~\ref{sec:results-one}, we investigate two examples: (i) particles whose propagation directions between subsequent active phases are anti-correlated by the CD with $-1/2\leq \kappa <0$ and (ii) particles showing an almost zero-turn event by the WC with $\kappa \rightarrow 1$.

\subsection{\label{sec:formalism-two}Two-state RTPs}

In the second scenario, we consider transport phenomena in a quasi-$2$D geometry [see Fig.~\ref{fig:formalism}(a)], where particles encounter separable dynamics modes, namely active and passive states (indicated by $a$ and $p$, respectively).
We formulate this two-state model by applying our one-state dynamics and local diffusive motion to $a$ and $p$, respectively, and concatenating these dynamic phases in a Markovian manner.
In detail, upon every completion of the ballistic motion, particles experience an overdamped Brownian motion with diffusivity $D$ for a duration $t_{2}$ drawn from an exponential distribution $\psi_{p}(t)=\gamma_{p}\exp(-\gamma_{p}t)$.
Note that the state transition satisfies the memoryless property via the Markovian embedding; however, the particles behave as if they memorize the direction of the previous run since we apply the identical $\theta$ dynamics described in Sec.~\ref{sec:formalism-one}.

We extend Eqs.~\eqref{eq:one-zeta} and \eqref{eq:one-prt} for our two-state model by successively utilizing the space-time coupled PDF on both states.
Let the particle reside in $\mathbf{r}-\mathbf{r}_{1}-\mathbf{r}_{2}$ at time $t-t_{1}-t_{2}$ with the velocity orientation $\theta-\phi$ [see Fig.~\ref{fig:formalism}(c)].
If turning angle $\phi$ and interarrival time $t_{1}$ are given, the particle shows a linear motion in the $\theta$ direction with a distance $vt_{1}$, following the PDF $p_{a}(\mathbf{r}_{1}|\theta,t_{1})=\delta (|\mathbf{r}_{1}|-vt_{1})\delta (\theta_{1}-\theta)/(vt_{1})$.
After reaching $\mathbf{r}-\mathbf{r}_{2}$ at time $t-t_{2}$, the particle immediately performs the Brownian motion for duration $t_{2}$, satisfying a PDF $p_{p}(\mathbf{r}_{2}|t_{2})=\exp[-\mathbf{r}_{2}^2/(4Dt_{2})]/(4\pi Dt_{2})$.
Eventually, the particle arrives at $\mathbf{r}$ strictly at time $t$.
Thus, we determine
\begin{widetext}
\begin{align}
    \tilde{\zeta} (\mathbf{r}, \theta, t) =
    \int_{0}^{t}dt_{2}
    \int_{\mathbb{R}^{2}} d^{2}\mathbf{r}_{2} 
    \int_{0}^{t-t_{2}} dt_{1}
    \int_{-\pi}^{\pi} d\phi 
    \int_{\mathbb{R}^{2}}
    &\tilde{\zeta}(\mathbf{r}-\mathbf{r}_{1}-\mathbf{r}_{2}, \theta - \phi, t-t_{1}-t_{2}) \nonumber \\
    &\times Q(\phi) \psi_{a}(t_{1})p_{a}(\mathbf{r}_{1}|\theta,t_{1}) \psi_{p}({t}_{2}) p_{p}(\mathbf{r}_{2}|t_{2})  d^{2}\mathbf{r}_{1} 
    + \delta(\mathbf{r})\delta(t)P_{0}(\theta),
    \label{eq:two-zeta}
\end{align}
where $\tilde{\zeta} (\mathbf{r}, \theta, t)$ denotes the particle density of reaching $\mathbf{r}$ with the velocity orientation $\theta$ just completing the consecutive active and passive states at time $t$.
Also, we assign $P_{0}(\theta)=1/(2\pi)$ as the initial condition as in Sec.~\ref{sec:formalism-one}.
Here, the order of the state transition, e.g., whether $a\rightarrow p$ or $p\rightarrow a$ iteration, does not alter the PDF $\tilde{\zeta} (\mathbf{r}, \theta, t)$.

On the other hand, we underscore that the particles' states $i,j \in \{a, p\}$ affect the propagator $P_{ij}(\mathbf{r},t)$, i.e., the PDF of finding particles that start with initial state $i$ and occupy position $\mathbf{r}$ at time $t$ in finial state $j$.
Formally, we write
\begin{align}
    &P_{ii}(\mathbf{r},t) = 
    \int_{-\pi}^{\pi}d\theta 
    \int_{0}^{t}dt_{1} 
    \int_{-\pi}^{\pi}d\phi 
    \int_{\mathbb{R}^{2}} 
    \tilde{\zeta}(\mathbf{r}-\mathbf{r}_{1}, \theta - \phi, t-t_{1})
    Q(\phi) \Psi_{i}(t_{1}) p_{i}(\mathbf{r}_{1}|\theta,t_{1}) d^{2}\mathbf{r}_{1}, 
    \label{eq:two-prt-ii} \\
    &P_{ij}(\mathbf{r},t) = 
    \int_{-\pi}^{\pi}d\theta 
    \int_{0}^{t}dt_{2} 
    \int_{\mathbb{R}^{2}} d^{2}\mathbf{r}_{2} 
    \int_{0}^{t-t_{2}} dt_{1} 
    \int_{-\pi}^{\pi}d\phi 
    \int_{\mathbb{R}^{2}} 
    \tilde{\zeta}(\mathbf{r}-\mathbf{r}_{1}-\mathbf{r}_{2}, \theta - \phi, t-t_{1}-t_{2}) \nonumber \\
    &\hspace{8.3cm} \times Q(\phi) \psi_{i}({t}_{1}) p_{i}(\mathbf{r}_{1}|\theta,t_{1}) \Psi_{j}({t}_{2}) p_{j}(\mathbf{r}_{2}|\theta,t_{2}) d^{2}\mathbf{r}_{1},
    \label{eq:two-prt-ij}
\end{align}
\end{widetext}
where different state $i\neq j$ is assumed in Eq.~\eqref{eq:two-prt-ij}.
Here, the survival probability $\Psi_{i}(t)$ handles the uncompleted last step, and the marginalization is conducted via integrating over $\theta$.
Significantly, we have $p_{p}(\mathbf{r}|\theta,t)=p_{p}(\mathbf{r}|t)$, where $\theta$ acts as a virtual angle to hold the information for the subsequent active motion; hence, the short-range diffusion is irrelevant to $Q(\phi)$ in our two-state model.
Finally, we obatin $P_{i}(\mathbf{r},t)$, i.e., the PDF of finding particles with initial state $i$ at position $\mathbf{r}$ at time $t$, by summing over all final states $P_{i}(\mathbf{r},t) = \sum_{j\in \{a, p\}} P_{ij}(\mathbf{r},t)$.

We move Eq.~\eqref{eq:two-zeta} to the Fourier-Laplace domain and reach
\begin{equation}
    \hat{\tilde{\zeta}}(\mathbf{k},\theta, s)
    =\sum_{m,n=-\infty}^{\infty} \tilde{f}_{n}(\mathbf{k},s)\hat{\tilde{\zeta}}_{m} (\mathbf{k}, s) Q_{m}e^{i(m-n)\theta} +\ P_{0}(\theta)
    \label{eq:two-zeta-hat}
\end{equation}
using the expansion $\hat{\tilde{\zeta}}(\mathbf{k},\theta, s)=\sum_{n=-\infty}^{\infty}\hat{\tilde{\zeta}}_{n}(\mathbf{k},s)e^{in\theta}$.
Analogously to the one-state model, we encapsulate functions into
\begin{align}
    \tilde{f}_{n}(\mathbf{k},s)
    = (i)^{-n}e^{in\psi}
    \mathcal{L} \{ &\psi_{a}(t_{1})J_{n}(|\mathbf{k}| vt_{1}) \}
    \nonumber \\
    &\times \mathcal{L} \{ \psi_{p}(t_{2})\exp(-D\mathbf{k}^{2}t_{2}) \},
    \label{eq:two-tilde-fn}
\end{align}
which satisfies $\tilde{f}_{n}(\mathbf{k},s)=f_{n}(\mathbf{k},s)\mathcal{L} \{ \psi_{p}(t_{2})\exp(-D\mathbf{k}^{2}t_{2}) \}$ [see Eq.~\eqref{app:eq:two-tilde-fn} for explicit expression].

Equation~\eqref{eq:two-zeta-hat} gives the truncated matrix equation for a non-negative cutoff index $\mu$:
\begin{equation}
    (I-\tilde{\mathcal{A}}\mathcal{D})\tilde{Z}^{\mu} = E_{0},
    \label{eq:two-matrix-equation}
\end{equation}
where the column vector $\tilde{Z}^{\mu}$ satisfies $\lim_{\mu\rightarrow \infty} \tilde{Z}_{\nu}^{\mu}(\mathbf{k},s)=\hat{\tilde{\zeta}}_{\nu}(\mathbf{k},s)$ for each Fourier mode index $\nu$ where $-\mu\leq \nu\leq \mu$.
Also, we find a modified Toeplitz matrix $\tilde{\mathcal{A}}=[\tilde{\mathcal{A}}_{mn}]$ with the component $\tilde{\mathcal{A}}_{mn}=\tilde{f}_{n-m}(\mathbf{k},s)$ for indices $m,n \in \left \{1, \dots, 2\mu+1 \right \}$.
Due to the initial condition $P_{0}(\theta)=1/(2\pi)$, we have $E_{0}=[0, \dots, 1/(2\pi), \dots, 0]^{\mathrm{T}}$ identical to the one-state RTPs.
In the following section, we compute $\tilde{Z}^{\mu}_{0}$ in Eq.~\eqref{eq:two-matrix-equation} using the inverse of a matrix $\tilde{\mathcal{B}}:=I-\tilde{\mathcal{A}}\mathcal{D}$.

We perform the Fourier-Laplace transform of Eqs.~\eqref{eq:two-prt-ii} and \eqref{eq:two-prt-ij} and consequently yield $\hat{P}_{ij}(\mathbf{k},s)$ for all possible states $i,j\in\{a,p\}$ [see Eqs.~\eqref{app:eq:two-pks-aj} and \eqref{app:eq:two-pks-pj}].
After marginalizing out the final states $j$, we obtain
\begin{equation}
    \hat{P}_{a}^{(\mu)}(\mathbf{k},s) =
    \big[
        \frac{2\pi}{\gamma_{p}} + \frac{2\pi(s+D\mathbf{k}^2+\gamma_{p})}{\gamma_{a}\gamma_{p}}
    \big] \big(
        \tilde{Z}_{0}^{\mu}-\frac{1}{2\pi}
    \big), \label{eq:two-final-pks-a}
\end{equation}
\begin{equation}
    \hat{P}_{p}^{(\mu)}(\mathbf{k},s) = 
    \big(
        \frac{2\pi}{\gamma_{a}}+\frac{2\pi}{s+D\mathbf{k}^2+\gamma_{p}}
    \big) \tilde{Z}_{0}^{\mu} - \frac{1}{\gamma_{a}}
    \label{eq:two-final-pks-p}
\end{equation}
for a non-negative integer $\mu$.
Here, the exponential property $\Psi_{i}(t)=\psi_{i}(t)/\gamma_{i}$ and normalization condition $Q_{0}=1$ are used.
In general, $\lim_{\mu \rightarrow \infty} \hat{P}^{(\mu)}_{i} (\mathbf{k}, s)=\hat{P}_{i} (\mathbf{k}, s)$ is fulfilled.

Let us remark on the dynamics at equilibrium.
Since our two-state model is a type of the Poissonian RTP, we follow a stationary solution of the Chapman-Kolmogorov equation (CKE) \cite{ross1995stochastic}: $P(\mathbf{r},t) = \sum_{i\in\{a,p\}}\pi_{i}P_{i}(\mathbf{r},t)$, where we define $\pi_{i}:=\tau_{i}/(\tau_{a}+\tau_{p})$ with $\tau_{i}=1/\gamma_{i}$ for each state $i \in \{a,p\}$.
The weight $\pi_{i}$ indicates the population of particles in state $i$ at a sufficiently large time $t$.

Lastly, we compute the $2\mu$th moment of displacement for the given initial state $i$ through
\begin{equation}
    \left \langle \mathbf{r}^{2\mu}(s) \right \rangle 
    = (-i)^{2\mu}\nabla_{\mathbf{k}}^{2\mu} \hat{P}_{i}^{(\mu)}(\mathbf{k},s)\big|_{\mathbf{k}=0},
    \label{eq:two-r2mu-del}
\end{equation}
whose Laplace inversion yields Eq.~\eqref{eq:one-r2mu-qmu-relation}.
Analogously to the one-state model, we can precisely obtain $\left \langle \mathbf{r}^{2\mu}(t) \right \rangle$ utilizing the truncated PDFs in Eqs.~\eqref{eq:two-final-pks-a} and \eqref{eq:two-final-pks-p}, which require only the Fourier coefficients $Q_{\nu}$ for $|\nu| \leq \mu$, not the entire $Q(\phi)$ (see also Appendix~\ref{app:sec:proof}).

\section{\label{sec:results-one}Results For One-state RTPs}

\subsection{\label{sec:results-one-a}Position distribution}

We investigate the Van Hove function $P(\mathbf{r},t)$ of the one-state RTPs.
First, when $Q(\phi)$ is the CD, we obtain a $3$\nobreakdash-by\nobreakdash-$3$ matrix $\mathcal{D}=\mathrm{diag}(\alpha_{1}, 1, \alpha_{1})$ in Eq.~\eqref{eq:one-matrix-equation}.
Accordingly, $\hat{P}(\mathbf{k},s)$ is equivalent to $\hat{P}^{(1)}(\mathbf{k},s)$ and Eq.~\eqref{eq:one-pks-mu} yields
\begin{align}
    \hat{P}(\mathbf{k},s)=
    \frac{
        \mathbf{k}^2v^2
        +4\gamma_{a}\alpha_{1}(\gamma_{a}(1-\alpha_{1})+s)
    }{
        \splitfrac{\textstyle  
            \mathbf{k}^2v^2 \left [
                \gamma_{a}(2\alpha_{1}-1)
                + \sqrt{(\gamma_{a}+s)^2+\mathbf{k}^2v^2}
                \right ]
        }{\textstyle
                + 4\gamma_{a}\alpha_{1} s(\gamma_{a}(1-\alpha_{1})+s) \hspace{-1.65mm}
        }
    }
    \label{eq:results-one-cd-pks}
\end{align}
for $\left | \alpha_{1} \right | \leq 1/2$.
We observe Eq.~\eqref{eq:results-one-cd-pks} is statistically isotropic since the function depends only on $|\mathbf{k}|$.
For $\alpha_{1}=0$, the CD reduces to the uniform distribution and Eq.~\eqref{eq:results-one-cd-pks} becomes Eq.~\eqref{app:eq:convention-pks}, i.e., $\hat{P}(\mathbf{k},s)$ of the conventional RTPs.
Note that its Fourier-Laplace inversion $P(\mathbf{r},t)$ is already known \cite{martens2012probability, santra2020run} [see Eq.~\eqref{app:eq:convention-pxt}].
On the other hand, to examine the non-uniform case $\alpha_{1}\neq 0$, we take the large $(\mathbf{k}, s)$ and the small $(\mathbf{k}, s)$ limits and get the asymptotic form of Eq.~\eqref{eq:results-one-cd-pks} in the short- and long-time regimes, respectively.

\begin{figure}[!t]
\centering
\includegraphics[keepaspectratio=true, width=8.6cm, height=100cm]{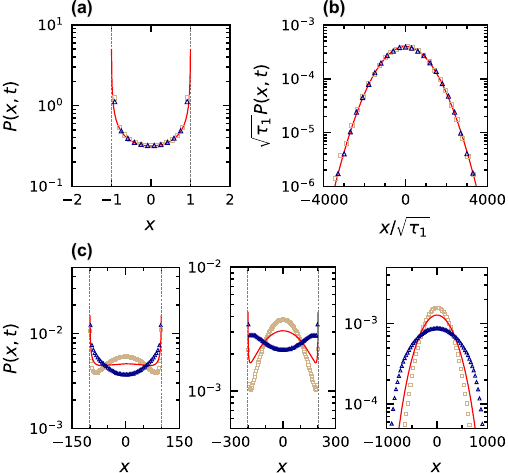}  
\caption{
Marginalized PDFs $P(x,t)$ for fixed $\gamma_{a}=0.01$ and $v=1$ with different times: (a) $t=1$, (b) $10^4$, (c) $100$, $200$, and $1000$ (left to right).
The red solid line represents the PDF of the conventional RTPs utilizing $Q(\phi)=1/(2\pi)$ [Eq.~\eqref{app:eq:convention-pxt}], whereas the markers are obtained from Monte Carlo simulations when $Q(\phi)$ is the CD with $\alpha_{1}=-1/2$ (yellow) and $1/2$ (blue).
We generated $N=2\times10^5$ realizations for each case.
(a) $P(x,t)$ at short times.
Note that since the propagator does not depend on $Q(\phi)$, all cases perfectly collapse to Eq.~\eqref{eq:results-one-short-pxt}.
(b) $\sqrt{\tau_{1}}P(x,t)$ versus rescaled variable $x/\sqrt{\tau_{1}}$ at long times, following Eq.~\eqref{eq:results-one-long-pxt}.
The variance of the Gaussian PDF widens as $\tau_{1}$ grows.
(c) $P(x,t)$ at intermediate times.
The higher angular moments $\alpha_{\mu\geq2}$ affect the PDF only at this intermediate-time regime.
}
\label{fig:results-one-pdf}
\end{figure}

Let us consider the short-time behavior of Eq.~\eqref{eq:results-one-cd-pks}.
In $s$ fixed and $\mathbf{k}\rightarrow \infty$ limit, we retain the highest order terms in $\mathbf{k}$ in the numerator and denominator, yielding $\hat{P}(\mathbf{k},s) \approx 1/\sqrt{(\gamma_a+s)^2+\mathbf{k}^2v^2}$.
Owing to its rotational symmetry in the Fourier domain, we substitute $k_{y}=0$ and perform the inverse Fourier transform about the $k_{x}$\nobreakdash-axis.
Then, we obtain $\mathcal{L}\{P(x,t)\}(x,s)=1/(\pi v)K_{0}[(\gamma_{a}+s)x/v]$, where $K_{0}$ is the modified Bessel function of the second kind \cite{bateman1954tables}.
Exploiting the shift theorem of the Laplace transform, we implement the Laplace inversion and obtain
\begin{equation}
    P(x,t) \approx  \frac{1}{\pi}\frac{1}{\sqrt{v^2t^2-x^2}} \exp \big [ 
        -\gamma_{a}t
    \big ],
    \label{eq:results-one-short-pxt}
\end{equation}
which is the well-known arcsinus law \cite{feller1991introduction, klafter2011first, froemberg2015asymptotic} modulated by an exponential relaxation $\exp(-\gamma_{a}t)$ \cite{santra2020run}.
Here, Eq.~\eqref{eq:results-one-short-pxt} does not depend on $\alpha_{1}$, so the first turn event can hardly occur at short times.
It is consistent with simulation results (yellow and blue) in Fig.~\ref{fig:results-one-pdf}(a).

At the long-time limit, $\sqrt{(\gamma_a+s)^2+\mathbf{k}^2v^2} \approx  (\gamma_a+s)$ is inserted in Eq.~\eqref{eq:results-one-cd-pks}.
By retaining the lowest order terms in $s$ and $\mathbf{k}$, we find $\hat{P}(\mathbf{k},s) \approx 2\gamma_{a}^{2}\alpha_{1}(1-\alpha_{1})/[\gamma_{a}\alpha_{1} \mathbf{k}^{2}v^{2}+2\gamma_{a}^{2}\alpha_{1}(1-\alpha_{1})s]$.
After performing the Fourier-Laplace inversion and marginalization $P(x,t)=\int_{-\infty}^{\infty}P(\mathbf{r},t)dy$,
\begin{equation}
    P(x,t) \approx  
    \frac{1}{\sqrt{2\pi v^2\tau_{1}t}} \exp \big [ 
        -\frac{x^2}{2v^2\tau_{1}t} \big ]
    \label{eq:results-one-long-pxt}
\end{equation}
is recovered, where the characteristic timescale $\tau_{1}$ is given as
\begin{equation}
    \tau_{1}
    = \frac{1}{\gamma_{a}(1-\alpha_{1})}.
    \label{eq:tau1}
\end{equation}
We plot $\sqrt{\tau_{1}}P(x,t)$ versus rescaled variable $x/\sqrt{\tau_{1}}$ for fixed $\gamma_{a}=0.01$ and $v=1$ in Fig.~\ref{fig:results-one-pdf}(b).
We detect a perfect overlap for various $\alpha_{1}$; in other words, when $Q(\phi)$ has larger $\alpha_{1}$, the modified duration $\tau_{1}$ also becomes longer, resulting in the growing variance of $P(x,t)$.

At the intermediate-time regime, we verify $\hat{P}(\mathbf{k}, s)$ by numerically performing the inverse Fourier-Laplace transform of Eq.~\eqref{eq:results-one-cd-pks} [see Fig.~\ref{app:fig:simulations}(a)].
Here, we plot the corresponding Monte Carlo simulation results in Fig.~\ref{fig:results-one-pdf}(c).
For all cases, the central part of the three PDFs rises to recover the Gaussian distribution as time passes.
When compared to the conventional RTPs (red), the system with $\alpha_{1}=-1/2$ (yellow) evolves faster, while $\alpha_{1}=1/2$ (blue) grows slower.
In Sec.~\ref{sec:results-one-c}, we quantitatively evaluate how the cosine moments $\alpha_{1}$ and $\alpha_{2}$ regulate this Gaussian recovery by deriving a non-Gaussianity $\mathrm{NG}(t)$ at all $t$.

Next, when $Q(\phi)$ is the WC, we confront an infinite-sized $\mathcal{D}$ in Eq.~\eqref{eq:one-matrix-equation} because of its non-vanishing cosine moments, which leads it challenging to find $\hat{P}(\mathbf{k},s)$ in a closed form.
Nevertheless, we can corroborate the short-time PDF to be Eq.~\eqref{eq:results-one-short-pxt} since the tumbling event that depends on $Q(\phi)$ is unlikely extant for such narrow time windows.
For the long-time PDF, we again exploit Eq.~\eqref{eq:results-one-long-pxt} restricting $0\leq \alpha_{1}<1$, because the CLT must be fulfilled regardless of $Q(\phi)$.
Consequently, the difference of $P(x,t)$ appears only at intermediate times.
To quantify this difference, we calculate the time-dependent cosine moment $\mathcal{M}_{\mu}(t)$ of $Q(\phi)$ in Sec.~\ref{sec:results-one-d}.

\subsection{\label{sec:results-one-b}Mean-squared displacement}

We can directly calculate the MSD using Eq.~\eqref{eq:results-one-cd-pks}.
We obtain $\left \langle \mathbf{r}^{2}(s) \right \rangle$ from Eq.~\eqref{eq:one-r2mu-del} with $\mu=1$.
After performing the inverse Laplace transform, we find
\begin{equation}
    \left \langle \mathbf{r}^{2}(t) \right \rangle = 
    \frac{2v^{2}}{\gamma_{a}(1-\alpha_{1})}t
    +\frac{2v^{2}}{\gamma_{a}^{2}(1-\alpha_{1})^{2}}
    (e^{-\gamma_{a}(1-\alpha_{1})t}-1)
    \label{eq:results-one-msd}
\end{equation}
consistent with Refs.~\cite{taktikos2013motility, villa2020run, sevilla2020two}.
Note that regardless of the $Q(\phi)$ type, we derive the identical MSD as long as $\alpha_{1}$ is kept fixed.
Also, when $\alpha_{1}$ approaches zero, Eq.~\eqref{eq:results-one-msd} recovers Eq.~\eqref{eq:rtp-convention-msd}, i.e., $\left \langle \mathbf{r}^{2}(t) \right \rangle$ of the conventional RTPs.
Comparing Eq.~\eqref{eq:results-one-msd} with $\left \langle \mathbf{r}^{2}(t) \right \rangle$ of the ABPs [Eq.~\eqref{eq:abp-msd}], the timescale $\tau_{1}$ replaces $1/D_{\mathrm{R}}$.

\begin{figure}[!t]
\centering
\includegraphics[keepaspectratio=true, width=8.6cm, height=100cm]{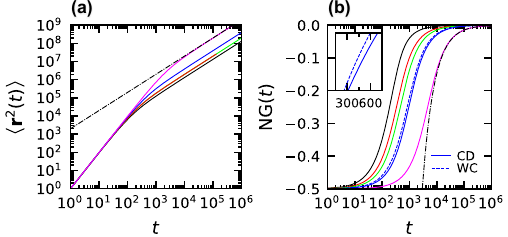}  
\caption{
MSDs $\left \langle \mathbf{r}^2(t) \right \rangle$ and non-Gaussianity $\mathrm{NG}(t)$ for different $\alpha_{1}=-0.5$ (black), $0$ (red), $0.5$ (blue), and $0.9$ (magenta) at fixed $\gamma_{a}=0.01$ and $v=1$; or $\gamma_{a}=0.1$, $v=1$, and $\alpha_{1}=0.9$ (green).
(a) $\left \langle \mathbf{r}^2(t) \right \rangle$ [Eq.~\eqref{eq:results-one-msd}].
All cases initially manifest the ballistic scaling of Eq.~\eqref{eq:results-one-short-r2}, and as $t\rightarrow\infty$, they recover the Fickian scaling of Eq.~\eqref{eq:results-one-long-r2} (dash-dotted line).
Here, the red and green curves are indistinguishable due to the identical $\tau_{1}=100$.
(b) $\mathrm{NG}(t)$ in semi-log scale [Eq.~\eqref{app:eq:one-ng}].
The parameters used are $\alpha_{2}=0$ from the CD (black, red, and blue solid lines) and $\alpha_{2}=\alpha_{1}^2$ from the WC (blue dashed line; and green and magenta solid lines).
As $t\rightarrow\infty$, $\mathrm{NG}(t)$ follows Eq.~\eqref{eq:ng-approximation-long} (dash-dotted line), converging to a plateau $\mathrm{NG}(t)=0$.
The two blue curves demonstrate a disparate time evolution due to distinct $\alpha_{2}$ values, as clearly shown in (b, inset).
}
\label{fig:results-one-msd-ng}
\end{figure}

We examine the asymptotic behavior of Eq.~\eqref{eq:results-one-msd}.
At the long-time limit, Eq.~\eqref{eq:results-one-msd} exhibits the Fickian scaling $\propto t$, strictly following
\begin{equation}
    \lim_{t\rightarrow \infty}\left \langle \mathbf{r}^2(t) \right \rangle = 4D_{\mathrm{eff}}t,
    \label{eq:results-one-long-r2}
\end{equation}
where the effective diffusivity $D_{\mathrm{eff}}$ is defined as
\begin{equation}
    D_{\mathrm{eff}}
    = \frac{v^2}{2\gamma_{a}(1-\alpha_{1})}.
    \label{eq:results-one-diffusivity}
\end{equation}
On the other hand, in the limit $t \rightarrow 0$, we obtain
\begin{equation}
    \left \langle \mathbf{r}^2(t) \right \rangle \approx  v^{2}t^{2}
    \label{eq:results-one-short-r2}
\end{equation}
independent of parameters $\gamma_{a}$ and $\alpha_{1}$.
It is consistent with the merged curves in Fig.~\ref{fig:results-one-msd-ng}(a) at short times.
We note that the equivalent limiting behavior is expected for the long-time approximation, e.g., in $t\rightarrow \infty, \gamma_{a}\rightarrow 0$ or $t\rightarrow \infty, \alpha_{1}\rightarrow 1$ limit, where the particle inevitably undergoes a zero-turn event.

We plot the MSD [Eq.~\eqref{eq:results-one-msd}] for various $\tau_{1}$ at fixed $v=1$ in Fig.~\ref{fig:results-one-msd-ng}(a).
At short times, particles obey Eq.~\eqref{eq:results-one-short-r2} exhibiting the scaling $\propto t^{2}$ regardless of $\tau_{1}$.
As time passes, they follow Eq.~\eqref{eq:results-one-long-r2}, which possesses $D_{\mathrm{eff}}$ that depends on $\tau_{1}$.
In particular, the particles simulated by the CD with $\alpha_{1}=-0.5$ (black) undergo anti-correlated reorientation during the tumble and hence display a narrow dispersion ($D_{\mathrm{eff}}$) compared to the conventional RTPs (red), which is precisely opposite to the case with $\alpha_{1}=0.9$ (magenta).
More significantly, focusing on the cases of $(\gamma_{a},\alpha_{1})=(0.01,0)$ (red) and $(0.1,0.9)$ (green), even though their $Q(\phi)$ is different, we cannot distinguish their dynamics due to the identical $\tau_{1}=100$.
Thus, we need to calculate the higher moments of displacement, which demand $\alpha_{\mu\geq 2}$ and cannot be scaled with a single timescale $\tau_{1}$.

\subsection{\label{sec:results-one-c}Non-Gaussianity}

The non-Gaussianity $\mathrm{NG}(t)$ in the $2$D plane is given as
\begin{equation}
    \mathrm{NG} (t) := \frac{
        \left \langle \mathbf{r}^4(t) \right \rangle
    }{
        3/2 \left \langle \mathbf{r}^2(t) \right \rangle^2  
    } - 1.
    \label{eq:ng-definition}
\end{equation}
To attain the explicit expression of Eq.~\eqref{eq:ng-definition}, we find $\left \langle \mathbf{r}^4(t) \right \rangle$ \cite{villa2020run, sevilla2020two} by utilizing $Z_{0}^{2}$ that is derivable from Eqs.~\eqref{eq:one-matrix-equation}, \eqref{eq:one-pks-mu}, and \eqref{eq:one-r2mu-del} with $\mu=2$ [see Eq.~\eqref{app:eq:one-r4}].
By the definition, we obtain $\mathrm{NG}(t)$, whose full expression is given in Eq.~\eqref{app:eq:one-ng}.
We note that $\mathrm{NG}(t)$ is a nondimensionalized fourth moment of displacement and does not depend on the speed $v$.

Let us demonstrate the limiting behavior of $\mathrm{NG}(t)$.
After we approximate the exponential terms up to the second order in $t$, we find $\mathrm{NG}(t) \approx -0.5$ at short times regardless of $Q(\phi)$.
This constant value originates from the PDF at short times where the contribution of $\alpha_{\mu}$ disappears.
At long times, we have
\begin{equation}
    \lim_{t\rightarrow \infty}\mathrm{NG} (t) = -\frac{2\tau_{1}-\tau_{2}}{t},
    \label{eq:ng-approximation-long}
\end{equation}
where we introduce the second characteristic timescale
\begin{equation}
    \tau_{2}
    = \frac{1}{\gamma_{a}(1-\alpha_{2})}.
    \label{eq:tau2}
\end{equation}
Equation~\eqref{eq:ng-approximation-long} demonstrates that $\tau_{2}$ determines the rate of Gaussian relaxation along with $\tau_{1}$ [Eq.~\eqref{eq:tau1}].

In Fig.~\ref{fig:results-one-msd-ng}(b), we plot $\mathrm{NG}(t)$ [Eq.~\eqref{app:eq:one-ng}] for the same parameters used in Fig.~\ref{fig:results-one-msd-ng}(a).
We first confirm that all particles exhibit the identical asymptotic behavior, $\lim_{t\rightarrow 0}\mathrm{NG}(t)= -0.5$ at short times and $\lim_{t\rightarrow \infty}\mathrm{NG}(t)= 0$ at long times.
Also, as expected from Eq.~\eqref{eq:ng-approximation-long}, we observe that if $|2\tau_{1}-\tau_{2}|$ becomes large, $\mathrm{NG}(t)$ more slowly recovers the Gaussianity.
For example, among the cases where the CD is characterized by $\alpha_{2}=0$ (black, red, and blue solid lines), the condition with smaller $\tau_{1}$, i.e., smaller $D_{\mathrm{eff}}$ here, brings out $\mathrm{NG} (t)=0$ more quickly.
Likewise, when comparing the cases of $(\gamma_{a},\alpha_{1})=(0.01,0)$ (red) and $(0.1,0.9)$ (green), the green line evolves more slowly even though both possess identical $\tau_{1}$.
Similarly, the inset in Fig.~\ref{fig:results-one-msd-ng}(b) shows that different $\alpha_{2}$ evokes the deviation in $\mathrm{NG}(t)$ even in cases with the equal $\alpha_{1}=0.5$ (blue solid and dashed lines).
Conclusively, when the other parameters $\gamma_{a}$ and $v$ are fixed, $\left \langle \mathbf{r}^4(t) \right \rangle$ or $\mathrm{NG}(t)$ is entirely characterized by $\alpha_{1}$ and $\alpha_{2}$ and is irrelevant to the higher cosine moments $\alpha_{\mu\geq 3}$.

\subsection{\label{sec:results-one-d}Time-dependent cosine moment}

Equation~\eqref{eq:one-r2mu-qmu-relation} states that the $2\mu$th moment of displacement $\left \langle \mathbf{r}^{2\mu}(t) \right \rangle$ depends on the finite cosine moments $\alpha_{\nu}$ for $|\nu|\leq \mu$.
Therefore, as a macroscopic observable that parameterizes the entire $Q(\phi)$, we suggest the angular distribution $\mathcal{Q}(\Theta , t)$ of an orientation variable $\Theta(t)$, i.e., the relative directional shift between $\mathbf{v}(0)$ and $\mathbf{v}(t)$, defined as
\begin{equation}
    \Theta (t)
    :=\arccos \big[
        \mathbf{v}(t)\cdot \mathbf{v}(0)/v^2
    \big].
    \label{eq:vel-orientation-definition}
\end{equation}
At time $t=0$, since we can regard all particles are traveling straight along the positive $x$\nobreakdash-axis with $\mathbf{v}=v\mathbf{\hat{x}}$, the governing equations for $\Theta (t)$ become Eqs.~\eqref{app:eq:one-vel-orientation-zeta} and \eqref{app:eq:one-vel-orientation-pdf}.
After solving the equations with the exponential function $\psi_{a}(t)$ and symmetric distribution $Q(\phi)$ (see Appendix~\ref{app:sec:derivation}), we reach
\begin{equation}
    \mathcal{Q}(\Theta  , t)
    =\frac{1}{2\pi} \big [ 
        1 + 2\sum_{n=1}^{\infty} \exp [
            -\gamma_{a}(1-\alpha_{n}) t
        ] \cos n\Theta
    \big ].
    \label{eq:results-one-vel-orientation-pdf}
\end{equation}
As time $t\rightarrow\infty$, it reduces to the uniform distribution $U(\Theta)=1/(2\pi)$ expected from the CLT.

\begin{figure}[!t]
\centering
\includegraphics[keepaspectratio=true, width=8.6cm, height=100cm]{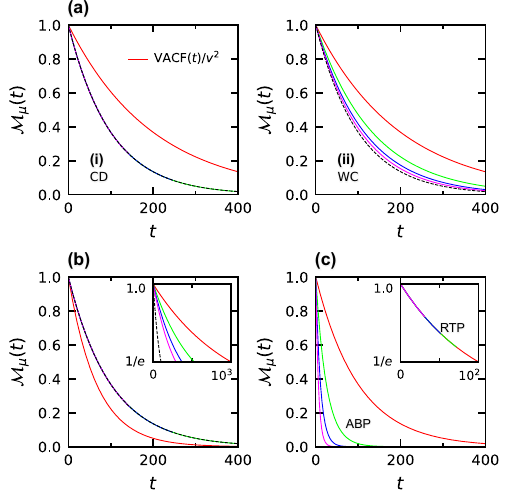}  
\caption{
Time-dependent cosine moment $\mathcal{M}_{\mu}(t)$ when the order $\mu=1$ (red), $2$ (green), $3$ (blue), and $4$ (magenta) for fixed $\gamma_{a}=0.01$ and $v=1$.
$\mathcal{M}_{1}(t)$ corresponds to the normalized $\mathrm{VACF}(t)$ (red).
(a), (b) RTPs with various $Q(\phi;\kappa)$.
The dashed line indicates an asymptotic function $\mathcal{M}_{\infty}(t)=\exp(-\gamma_{a}t)$.
(a) RTPs whose $Q(\phi;\kappa)$ is (i) the CD with $\kappa=0.5$ [Eq.~\eqref{eq:results-one-cosine-cd}] and (ii) WC with $\kappa=0.5$ [Eq.~\eqref{eq:results-one-cosine-wc}].
(b) RTPs whose $Q(\phi;\kappa)$ is the CD with $\kappa=-0.5$ and (b, inset) WC with $\kappa=0.9$.
We can distinguish (a)-(i) and (a)-(ii) through $\mathcal{M}_{\mu\geq2}(t)$.
(c) ABPs when $D_{\mathrm{R}}=\gamma_{a}$.
(c, inset) RTPs with the uniform distribution.
The angular moments of ABPs [Eq.~\eqref{eq:results-one-cosine-wn}] are sharply disbanded as $\mu$ increases, whereas the conventional RTPs (c, inset) possess a single curve $\mathcal{M}_{\mu}(t)=\exp(-\gamma_{a}t)$, 
See Fig.~\ref{app:fig:simulations}(d)--\ref{app:fig:simulations}(f) to check the agreement with the simulation results in Appendix~\ref{app:sec:simulations}.
}
\label{fig:results-one-cosine}
\end{figure}

We extract the $\mu$th time-dependent cosine moment $\mathcal{M}_{\mu}(t)$ from Eq.~\eqref{eq:results-one-vel-orientation-pdf}.
Following the definition of the cosine moment [Eq.~\eqref{eq:alpha-definition}], we obtain
\begin{equation}
    \mathcal{M}_{\mu}(t)
    =\exp[-\gamma_{a}(1-\alpha_{\mu})t]
    \label{eq:cosine-moment-definition}
\end{equation}
for a positive integer $\mu$.
Let us indicate the $\mu$th characteristic timescale as $\tau_{\mu}=1/[\gamma_{a}(1-\alpha_{\mu})]$.
Then, the maximum value $\max_{\mu}(\tau_{\mu})$ defines the crossover timescale over which $\mathcal{Q}(\Theta , t)$ can be approximated to $U(\Theta)=1/(2\pi)$.
More specifically, if $Q(\phi;\kappa)$ follows the CD, we have
\begin{equation}
    \mathcal{M}_{\mu}(t) =
    \begin{cases}
        \exp[-\gamma_{a}(1-\kappa)t], & \mu=1, \\ 
        \exp [ -\gamma_{a}t ], & \mu\geq2
    \end{cases}
    \label{eq:results-one-cosine-cd}
\end{equation}
for $|\kappa| \leq 1/2$ due to its vanishing $\alpha_{\mu\geq2}$.
For the WC, we get
\begin{equation}
    \mathcal{M}_{\mu}(t) = \exp [ -\gamma_{a}(1-\kappa^{\mu})t ]
    \label{eq:results-one-cosine-wc}
\end{equation}
for $0 \leq \kappa < 1$.
On the contrary, the simplest case with $Q(\phi)=1/(2\pi)$ (the conventional RTPs) results in a single curve $\mathcal{M}_{\mu}(t)=\exp (-\gamma_{a}t)$ regardless of the order $\mu$.
Next, we also address the cosine moments of the ABPs that are described by a wrapped normal (WN) distribution $Q(\phi;\sigma^{2})$ [Eq.~\eqref{app:eq:wrapped-normal}] \cite{romanczuk2012active}.
In this case, we yield
\begin{equation}
    \mathcal{M}_{\mu}(t)=\exp[-D_{\mathrm{R}}\mu^{2}t],
    \label{eq:results-one-cosine-wn}
\end{equation}
where we substitute $\sigma^2=2D_{\mathrm{R}}t$.
Note that the corresponding MSD is given in Eq.~\eqref{eq:abp-msd}.

We interpret the first cosine moment $\mathcal{M}_{1}(t)$ \cite{comments-kubo} in terms of the velocity autocorrelation function $\mathrm{VACF}(t)$ [Eq.~\eqref{app:eq:vacf-definition}].
By the definition of $\mathcal{M}_{1}(t)=\left \langle \cos \Theta(t)\right \rangle$, we directly obtain $\mathrm{VACF}(t)=v^2\exp[-\gamma_{a}(1-\alpha_{1})t]$ consistent with Ref.~\cite{taktikos2013motility}; that is, $\mathcal{M}_{1}(t)$ is the normalized $\mathrm{VACF}(t)$.
Particularly, we examine $\mathcal{M}_{1}(t)$ (red) for various SPPs at fixed $\gamma_{a}=0.01$ in Fig.~\ref{fig:results-one-cosine}.
For the one-state RTPs, $\alpha_{1}$ regulates the relaxation timescale $\tau_{1}$ to dissipate a directional persistence in active transport.
In other words, the RTPs with larger $\alpha_{1}$ maintain the positive two-point correlation (directionality) longer [see Figs.~\ref{fig:results-one-cosine}(a), \ref{fig:results-one-cosine}(b), and \ref{fig:results-one-cosine}(c, inset)].
However, $\mathcal{M}_{1}(t)$ (or $\mathrm{VACF}(t)$) cannot be a proper observable to distinguish RTPs with different $Q(\phi)$ if they possess the same $\alpha_{1}$ [see Figs.~\ref{fig:results-one-cosine}(a)-(i) (CD) and \ref{fig:results-one-cosine}(a)-(ii) (WC)].
Likewise, when we designate $D_{\mathrm{R}}=\gamma_{a}$ in Fig.~\ref{fig:results-one-cosine}(c), the ABPs and the conventional RTPs manifest indistinguishable velocity relaxation for the demonstrated timescale $\tau_{1}=100$.

Hence, we investigate the high-resolved stochastic quantity $\mathcal{M}_{\mu\geq2}(t)$ to capture the inherent characteristics of the RTP dynamics that originate from $Q(\phi)$.
Considering $\mathcal{M}_{\mu}(t)$ for the order $2\leq\mu\leq4$, we can now discriminate the RTPs with different $Q(\phi)$ despite the equal $\alpha_{1}=0.5$ in Fig.~\ref{fig:results-one-cosine}(a).
In particular, the curves with the CD [Fig.~\ref{fig:results-one-cosine}(a)-(i)] all overlay for $\mu\geq2$, whereas the curves with the WC [Fig.~\ref{fig:results-one-cosine}(a)-(ii)] are disbanded and gradually approach $\mathcal{M}_{\infty}(t)=\exp(-\gamma_{a}t)$ (dashed line).
Similarly, $\mathcal{M}_{\mu\geq2}(t)$ in Fig.~\ref{fig:results-one-cosine}(c) reveals the distinctive angle dynamics embedded in ABPs and conventional RTPs.
Conclusively, $\mathcal{Q}(\Theta, t)$ can unveil the turning patterns of RTPs more strictly than $\left \langle \mathbf{r}^{2\mu}(t) \right \rangle$.

\section{\label{sec:results-two}Results For Two-state RTPs}

\subsection{\label{sec:results-two-a}Position distribution}

We examine the Van Hove function $P_{i}(\mathbf{r},t)$ of the two-state RTPs for each initial state $i\in\{a,p\}$.
Let us first assume $Q(\phi)$ is the CD.
Then, we have a finite-sized $\mathcal{D}=\mathrm{diag}(\alpha_{1}, 1, \alpha_{1})$ in Eq.~\eqref{eq:two-matrix-equation}.
Since $\hat{P}_{i}^{(1)}(\mathbf{k},s)$ is equivalent to $\hat{P}_{i}(\mathbf{k},s)$, Eqs.~\eqref{eq:two-final-pks-a} and \eqref{eq:two-final-pks-p} determine $\hat{P}_{a}(\mathbf{k},s)$ [Eq.~\eqref{app:eq:two-cd-pks-a}] and $\hat{P}_{p}(\mathbf{k},s)$ [Eq.~\eqref{app:eq:two-cd-pks-p}], respectively.
Owing to their rotational symmetry with respect to $|\mathbf{k}|$, we focus on the PDFs marginalized over the $y$ direction, i.e., $P_{i}(x,t)$.

\begin{figure}[!t]
\centering
\includegraphics[keepaspectratio=true, width=8.6cm, height=100cm]{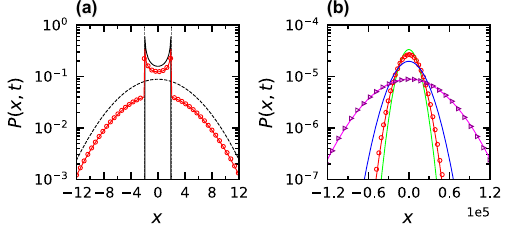}  
\caption{
Marginalized PDFs $P(x,t)$ for fixed $\gamma_{a}=0.01$, $\gamma_{p}=0.01$, $v=2$, and $D=10$ when (a) $t=1$ and (b) $10^{6}$.
The parameters used are $\alpha_{1}=-0.5$ (green), $0$ (red), $0.5$ (blue), and $0.9$ (magenta).
In Monte Carlo simulations, we generated $N=2\times10^{4}$ realizations.
(a) $P(x,t)$ at short times.
The propagator exhibits disparate behavior depending on the initial state $i\in\{a,p\}$.
In particular, when $i=a$, it follows $P_{a}(x,t)$ (black solid line); when $i=p$, it follows $P_{p}(x,t)$ (black dashed line).
At the equilibrium, the simulation markers (red circles) perfectly fit to Eq.~\eqref{eq:results-two-short-pxt} (red solid line).
(b) $P(x,t)$ at long times.
The markers (red and magenta) perfectly agree with Eq.~\eqref{eq:results-two-long-pxt}, which is valid for any symmetric distribution $Q(\phi)$.
}
\label{fig:results-two-pdf}
\end{figure}

First, we investigate the asymptotic form of $\hat{P}_{i}(\mathbf{k},s)$ in the short-time regime.
For the initial state $i=a$, Eq.~\eqref{app:eq:two-cd-pks-a} reduces to $\hat{P}_{a}(\mathbf{k},s) \approx  1/\sqrt{(\gamma_{a}+s)^{2}+\mathbf{k}^2 s^2}$ at the large $(\mathbf{k},s)$ limit.
Hence, Eq.~\eqref{eq:results-one-short-pxt} directly becomes its marginalized PDF $P_{a}(x,t)$.
On the other hand, for the initial state $i=p$, Eq.~\eqref{app:eq:two-cd-pks-p} shortens to $\hat{P}_{p}(\mathbf{k},s) \approx  1/(s + D\mathbf{k}^{2}+\gamma_{p})$ and we find $P_{p}(x,t) \approx  1/\sqrt{4\pi Dt}\exp[-x^2/(4Dt)-\gamma_{p}t]$.
At the equilibrium, we thus reach
\begin{align}
    P(x,t) \approx  
    &\pi_{a} \times \frac{1}{\pi}\frac{1}{\sqrt{v^2t^2-x^2}} \exp \big [ 
        -\gamma_{a}t
    \big ] \nonumber \\
    &+\pi_{p} \times \frac{1}{\sqrt{4\pi Dt}} \exp \big [ 
        -\frac{x^2}{4Dt} -\gamma_{p}t
    \big ],
    \label{eq:results-two-short-pxt}
\end{align}
where the first cosine moment $\alpha_{1}$ cannot affect the PDF.
In particular, we plot $P(x,t)$ [Eq.~\eqref{eq:results-two-short-pxt}] at $t=1$ for fixed $\gamma_{a}=\gamma_{p}=0.01$ in Fig.~\ref{fig:results-two-pdf}(a).
Here, we observe two peaks at $x=2$ and $-2$ due to the ballistic motion with $v=2$, reminiscent of L\'evy walk \cite{zaburdaev2015levy, klafter2011first}.
In contrast, at $|x|\gg vt$, it accompanies the Gaussian tails that stem from the local diffusion of the passive state.
We check that the theoretical curve (red line) displays excellent agreement with simulation results (red circles).

Next, we consider the long-time approximation of $\hat{P}_{i}(\mathbf{k},s)$.
Since the state transition obeys the stationary solution of the CKE at a sufficiently large time $t$, both $\hat{P}_{a}(\mathbf{k},s)$ and $\hat{P}_{p}(\mathbf{k},s)$ converge to $\hat{P}(\mathbf{k},s)$.
At the small $(\mathbf{k}, s)$ limit, we find $\hat{P}(\mathbf{k},s)$ [Eq.~\eqref{app:eq:two-long-pks}], and after performing the Fourier-Laplace inversion and marginalization, we reach
\begin{align}
    P(x,t) \approx  
    &\sqrt{
        \frac{
            \gamma_{a}(\gamma_{a}+\gamma_{p})(1-\alpha_{1})
        }{
            4\pi D\gamma_{a}^2(1-\alpha_{1})t + 2\pi\gamma_{p}v^{2}t
        }
    } \nonumber \\
    &\times
    \exp \left [ 
        -\frac{
            \gamma_{a}(\gamma_{a} + \gamma_{p})(1-\alpha_{1})x^2
        }{
            4D\gamma_{a}^2(1-\alpha_{1})t + 2\gamma_{p}v^{2}t
        }
    \right ]
    \label{eq:results-two-long-pxt}    
\end{align}
with $|\alpha_{1}|\leq1/2$.
As shown in Fig.~\ref{fig:results-two-pdf}(b), the first cosine moment $\alpha_{1}$ is responsible for controlling the variance of Eq.~\eqref{eq:results-two-long-pxt}.
In particular, as $\alpha_{1}$ becomes larger, e.g., from $\alpha_{1}=-1/2$ (green) to $1/2$ (blue), the dispersion of the function widens.

We note that Eq.~\eqref{eq:results-two-long-pxt} is valid for a general $Q(\phi)$ satisfying the reflective symmetry.
Thus, when $Q(\phi)$ is the WC, we still conclude Eq.~\eqref{eq:results-two-long-pxt} with $0\leq \alpha_{1}<1$ [see the agreement with simulation results in Fig.~\ref{fig:results-two-pdf}(b)].
In other words, the discrepancy of dynamics evoked by $\alpha_{\mu\geq2}$ is encoded only at the intermediate timescale.
Significantly, we also infer that two PDFs $P_{a}(x,t)$ and $P_{p}(x,t)$ undergo enormously distinct time evolution towards the identical Gaussian distribution [Eq.~\eqref{eq:results-two-long-pxt}].
Thus, in the following section, we characterize our two-state model by deriving the MSDs at all time regimes differing initial conditions: $i=a$, $p$, and equilibrium.

\subsection{\label{sec:results-two-b}Mean-squared displacement}

Analogously to the one-state RTPs in Sec.~\ref{sec:results-one-b}, we calculate the MSDs using Eqs.~\eqref{app:eq:two-cd-pks-a} and \eqref{app:eq:two-cd-pks-p}.
We first obtain $\left \langle \mathbf{r}^{2}(s) \right \rangle$ from Eq.~\eqref{eq:two-r2mu-del} with $\mu=1$ for each initial state $i\in\{a,p\}$.
After performing the inverse Laplace transform, we determine

\begin{widetext}
\begin{align}
    \left \langle \mathbf{r}^{2}(t) \right \rangle = 
    \frac{
        4D\gamma_{a}^2(1-\alpha_{1})+2v^2\gamma_{p}}{\gamma_{a}(1-\alpha_{1})(\gamma_{a}+\gamma_{p})
    }t
    -\frac{
        4D\gamma_{a}^3(1-\alpha_{1})^2
        -4v^{2}\gamma_{a}^{2}(1-\alpha_{1})
        +2v^{2}(\gamma_{a}+\gamma_{p})^{2}
        -2v^{2}\gamma_{a}\gamma_{p}(1-\alpha_{1})
    }{\gamma_{a}^2(1-\alpha_{1})^2(\gamma_{a}+\gamma_{p})^2}&
    \nonumber \\
    +\frac{2v^2e^{-(\gamma_{a}+\gamma_{p})t/2}}{
        \gamma_{a}^2(1-\alpha_{1})^2
    }
    \left[
        \frac{-\gamma_{a}^2+\gamma_{a}^2\alpha_{1}+2\gamma_{a}\gamma_{p}-3\gamma_{a}\gamma_{p}\alpha_{1}-\gamma_{p}^2}{\gamma_{p}\epsilon_{1}}\sinh(-\epsilon_{1} t/2)
        +\frac{-\gamma_{a}(1-\alpha_{1})+\gamma_{p}}{\gamma_{p}}\cosh(-\epsilon_{1} t/2) 
    \right]&
    \nonumber \\
    +\frac{
        4D\gamma_{a}\gamma_{p}(1-\alpha_{1})+2\gamma_{a}v^2}{\gamma_{p}(1-\alpha_{1})(\gamma_{a}+\gamma_{p})^2
    }e^{-(\gamma_{a}+\gamma_{p})t},&
    \label{eq:results-two-msd-a} \\
    \left \langle \mathbf{r}^{2}(t) \right \rangle =
    \frac{
        4D\gamma_{a}^2(1-\alpha_{1})+2v^2\gamma_{p}}{\gamma_{a}(1-\alpha_{1})(\gamma_{a}+\gamma_{p})
    }t
    +\frac{
        4D\gamma_{a}^2\gamma_{p}(1-\alpha_{1})^2
        +2v^{2}\gamma_{a}^{2}(1-\alpha_{1})
        -2v^{2}(\gamma_{a}+\gamma_{p})^2
    }{\gamma_{a}^2(1-\alpha_{1})^2(\gamma_{a}+\gamma_{p})^2}
    \hspace{3.5cm}&
    \nonumber \\
    +\frac{2v^2e^{-(\gamma_{a}+\gamma_{p})t/2}}{
        \gamma_{a}^2(1-\alpha_{1})^2
    }
    \left[
        \frac{\gamma_{a}-2\gamma_{a}\alpha_{1}-\gamma_{p}}{\epsilon_{1}}\sinh(-\epsilon_{1} t/2)
        +\cosh(-\epsilon_{1} t/2) 
    \right]
    -\frac{
        4D\gamma_{p}(1-\alpha_{1})+2v^2}{(\gamma_{a}+\gamma_{p})^2(1-\alpha_{1})
    }e^{-(\gamma_{a}+\gamma_{p})t},&
    \label{eq:results-two-msd-p}
\end{align}
where we set the initial state $i=a$ and $p$, respectively.
Let us denote $\left \langle \mathbf{r}_{i}^{2}(t) \right \rangle=\int_{\mathbb{R}^{2}} \mathbf{r}^{2}P_{i}(\mathbf{r},t) d^{2}\mathbf{r}$ as the MSD of particles initially in the state $i\in\{a,p\}$.
Then we obtain $\left \langle \mathbf{r}_{i}^{2}(t) \right \rangle=\sum_{j\in \{a, p\}}\int_{\mathbb{R}^{2}} \mathbf{r}^{2}P_{ij}(\mathbf{r},t) d^{2}\mathbf{r}=\sum_{j\in \{a, p\}}\pi_{j}\left \langle \mathbf{r}_{ij}^{2}(t) \right \rangle$, where $\left \langle \mathbf{r}_{ij}^{2}(t) \right \rangle$ indicates the MSD of particles initially in the state $i$ and observed in the state $j$ at time $t$.
We confirm that $\left \langle \mathbf{r}_{pa}^{2}(t) \right \rangle$ causes the hyperdiffusion at short times in Sec.~\ref{sec:results-two-with-rests}.
At the equilibrium, we follow the CKE, which leads to $\left \langle \mathbf{r}^{2}(t) \right \rangle=\sum_{i\in \{a, p\}} \pi_{i}\left \langle \mathbf{r}_{i}^{2}(t) \right \rangle$.
Consequently, we find
\begin{align}
    \left \langle \mathbf{r}^{2}(t) \right \rangle &=
    \frac{
        4D\gamma_{a}^2(1-\alpha_{1})+2v^2\gamma_{p}}{\gamma_{a}(1-\alpha_{1})(\gamma_{a}+\gamma_{p})
    }t
    -\frac{2v^{2}(\gamma_{a}\alpha_{1}+\gamma_{p})}{\gamma_{a}^2(1-\alpha_{1})^2(\gamma_{a}+\gamma_{p})}
    \nonumber \\
    &+\frac{2v^2e^{-(\gamma_{a}+\gamma_{p})t/2}}{
        \gamma_{a}^2(1-\alpha_{1})^2
    }
    \left[
        \frac{\gamma_{a}^2(1-\alpha_{1})+3\gamma_{a}\gamma_{p}(1-\alpha_{1})-(\gamma_{a}+\gamma_{p})^{2}}{(\gamma_{a}+\gamma_{p})\epsilon_{1}}\sinh(-\epsilon_{1} t/2)
        +\frac{\gamma_{a}\alpha_{1}+\gamma_{p}}{\gamma_{a}+\gamma_{p}}\cosh(-\epsilon_{1} t/2) 
    \right].
    \label{eq:results-two-msd-eq}
\end{align}
\end{widetext}
Throughout Eqs.~\eqref{eq:results-two-msd-a}--\eqref{eq:results-two-msd-eq}, we introduce a new rate parameter
\begin{equation}
    \epsilon_{1}=\sqrt{(\gamma_{a}+\gamma_{p})^{2}-4\gamma_{a}\gamma_{p}(1-\alpha_{1})}.
    \label{eq:epsilon1}
\end{equation}
Note that regardless of the $Q(\phi)$ type, we obtain the identical MSDs above, as long as $\alpha_{1}$ is equal.
Furthermore, since the Poissonian RTP guarantees the ergodicity, we can equate the ensemble-averaged MSD at the equilibrium [Eq.~\eqref{eq:results-two-msd-eq}] to a time-averaged MSD (TAMSD) $\overline{\delta^2(t)} := \int_{0}^{T-t} [\mathbf{r}(\tau+t)-\mathbf{r}(\tau)]^2d\tau /(T-t)$ at a sufficiently large $T$, where $T$ denotes a total measurement time.

\begin{figure}[!t]
\centering
\includegraphics[keepaspectratio=true, width=8.6cm, height=100cm]{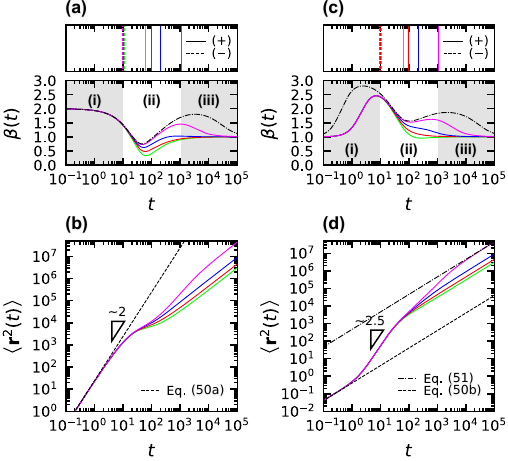}  
\caption{
Anomalous exponent $\beta(t)$ and MSDs $\left \langle \mathbf{r}^2(t) \right \rangle$ against time $t$ when (a), (b) initial state $i=a$ and (c), (d) $i=p$ for fixed $\gamma_{a}=0.1$ and $\gamma_{p}=0.01$.
The parameters used are $\alpha_{1}=-0.5$ (green), $0$ (red), $0.5$ (blue), and $0.9$ (magenta).
(a), (c) (top) Characteristic timescales $\tilde{\tau}_{1}^{(+)}$ (solid line) and $\tilde{\tau}_{1}^{(-)}$ (dashed line), following Eq.~\eqref{eq:tilde-tau1-pm}.
(bottom) $\beta(t)$ when $v=5$ and $D=0.1$ (solid line); or $v=25$, $D=0.1$, and $\alpha_{1}=0.99$ (dash-dotted line).
We divide the time regimes into the following three cases: (i) short times $t\lesssim\tilde{\tau}_{1}^{(-)}$, (ii) intermediate times $\tilde{\tau}_{1}^{(-)}\lesssim t\lesssim\tilde{\tau}_{1}^{(+)}$, and (iii) long times $t\gtrsim\tilde{\tau}_{1}^{(+)}$.
The shaded regions (gray) are divided based on $\alpha_{1}=0.9$ (magenta).
Note that the hyperdiffusion intensifies at a smaller $D/v^2$, as shown in the dash-dotted line in (c)-(i).
(b), (d) $\left \langle \mathbf{r}^2(t) \right \rangle$.
Regardless of the initial state $i$, MSDs eventually recover the Fickian scaling $\propto t$ given in Eq.~\eqref{eq:results-two-long-r2} (dash-dotted line).
}
\label{fig:results-two-msd}
\end{figure}

We investigate the asymptotic forms of Eqs.~\eqref{eq:results-two-msd-a}--\eqref{eq:results-two-msd-eq}.
First, in the limit $t \rightarrow 0$, we approximate Eqs.~\eqref{eq:results-two-msd-a} and \eqref{eq:results-two-msd-p} up to the lowest order term in $t$, resulting in
\begin{subnumcases}{
    \lim_{t\rightarrow 0}\left \langle \mathbf{r}^{2}(t) \right \rangle =
}
    (2D\gamma_{a}+v^{2}) t^{2}, & $i=a$, \label{eq:results-two-short-r2-a}\\
    4Dt, & $i=p$, \label{eq:results-two-short-r2-p}
\end{subnumcases}
respectively.
We again check that $Q(\phi)$ does not emerge as the control parameter at this short timescale.
On the contrary, in the limit $t \rightarrow \infty$, all MSDs from the different initial states increase linearly in time, precisely following
\begin{equation}
    \lim_{t\rightarrow \infty}\left \langle \mathbf{r}^2(t) \right \rangle = 4\tilde{D}_{\mathrm{eff}}t,
    \label{eq:results-two-long-r2}
\end{equation}
where the effective diffusivity $\tilde{D}_{\mathrm{eff}}$ is defined as
\begin{equation}
    \tilde{D}_{\mathrm{eff}}
    = \frac{2D\gamma_{a}^2(1-\alpha_{1})+v^2\gamma_{p}}{2\gamma_{a}(1-\alpha_{1})(\gamma_{a}+\gamma_{p})}.
    \label{eq:results-two-diffusivity}
\end{equation}
As expected from the CLT and Poissonian property of the two-state RTPs, the Fickian scaling $\propto t$ is recovered.
On the other hand, in the limit $t\rightarrow \infty$ and $\alpha_{1}\rightarrow 1$, we reach
\begin{equation}
    \left \langle \mathbf{r}^2(t) \right \rangle
    \approx v^2 \pi_{a}^2 t^{2}
    \label{eq:results-two-long-r2-zero-turn}
\end{equation}
for all MSDs in Eqs.~\eqref{eq:results-two-msd-a}--\eqref{eq:results-two-msd-eq}.
Because the condition $\alpha_{1} \rightarrow 1$ forces the particles at the active state to sustain their linear motion, $\left \langle \mathbf{r}^2(t) \right \rangle \approx  v^2t^2$ dominates as time passes, additionally multiplied by a factor of $\pi_{a}^{2}$ to solely pick the particles in the ballistic mode.
In other words, $\alpha_{1}$ manipulates the dynamics only when the particle resides in the active state.

Now, we focus on the hyperbolic terms in Eqs.~\eqref{eq:results-two-msd-a}--\eqref{eq:results-two-msd-eq}, where the detailed time evolution of $P_{i}(x,t)$ at intermediate times is encoded.
We find new characteristic timescales $\tilde{\tau}_{1}^{(+)}$ and $\tilde{\tau}_{1}^{(-)}$, given as
\begin{equation}
    \tilde{\tau}_{1}^{(\pm)}
    =\frac{2}{\gamma_{a}+\gamma_{p}\mp\epsilon_{1}}
    =\frac{\gamma_{a}+\gamma_{p}\pm\epsilon_{1}}{2\gamma_{a}\gamma_{p}(1-\alpha_{1})}.
    \label{eq:tilde-tau1-pm}
\end{equation}
We interpret $\tilde{\tau}_{1}^{(\pm)}$ as the modified mean duration for completing each state when $\alpha_{1}\neq0$.
As $\alpha_{1}$ approaches zero, we thus recover the means of $\psi_{a}(t)$ and $\psi_{p}(t)$, i.e., $\tilde{\tau}_{1}^{(+)}=1/\gamma_{<}$ and $\tilde{\tau}_{1}^{(-)}=1/\gamma_{>}$, where we denote $\gamma_{>}=\mathrm{max}(\gamma_{a}, \gamma_{p})$ and $\gamma_{<}=\mathrm{min}(\gamma_{a}, \gamma_{p})$.
Moreover, we evaluate a mean period to finish both active and passive states via $\tilde{\tau}_{1}:=\sum_{i\in\{+,-\}}\tilde{\tau}_{1}^{(i)}=(\gamma_{a}+\gamma_{p})/[\gamma_{a}\gamma_{p}(1-\alpha_{1})]$.
In this case, as $\alpha_{1} \rightarrow 0$, we attain $\tilde{\tau}_{1} = (\gamma_{a}+\gamma_{p})/(\gamma_{a}\gamma_{p})$, i.e., the mean of the hypoexponential distribution $\tilde{\psi}(t)=\gamma_{a}\gamma_{p}/(\gamma_{p}-\gamma_{a})[\exp(-\gamma_{a}t)-\exp(-\gamma_{p}t)]$.
More generally, we infer that $\left \langle \mathbf{r}^{2\mu}(t) \right \rangle $ involves the $\mu$th characteristic timescale $\tilde{\tau}_{\mu}^{(\pm)}=2/(\gamma_{a}+\gamma_{p}\mp\epsilon_{\mu})$, where we introduce $\epsilon_{\mu}=\sqrt{(\gamma_{a}+\gamma_{p})^{2}-4\gamma_{a}\gamma_{p}(1-\alpha_{\mu})}$ and $\tilde{\tau}_{\mu}=(\gamma_{a}+\gamma_{p})/[\gamma_{a}\gamma_{p}(1-\alpha_{\mu})]$ for a positive integer $\mu$.

We further explore the MSDs expressed in Eqs.~\eqref{eq:results-two-msd-a} and \eqref{eq:results-two-msd-p} for fixed $\gamma_{a}=0.1$ and $\gamma_{p}=0.01$ in Fig.~\ref{fig:results-two-msd}.
Here, we define an anomalous exponent $\beta(t):=\mathrm{dlog}\left \langle \mathbf{r}^{2}(t) \right \rangle/\mathrm{dlog}(t)$ to demonstrate the non-linear growth of MSDs.
We first observe the dissimilar time evolution of MSDs depending on whether the initial state is active or passive.
Specifically, for the initial state $i=a$ [Figs.~\ref{fig:results-two-msd}(a) and \ref{fig:results-two-msd}(b)], $\left \langle \mathbf{r}^{2}(t) \right \rangle$ displays a crossover from the ballistic motion $\lim_{t\rightarrow0}\beta(t)=2$ [Eq.~\eqref{eq:results-two-short-r2-a}] (dashed line) to the Fickian diffusion $\lim_{t\rightarrow\infty}\beta(t)=1$ [Eq.~\eqref{eq:results-two-long-r2}].
At intermediate times [Fig.~\ref{fig:results-two-msd}(a)-(ii)], $\beta(t)$ exhibits a drastic increase, for instance, from sub to superdiffusion at $\alpha_{1}=0.9$ (magenta).
Contrastingly, for the initial state $i=p$ [Figs.~\ref{fig:results-two-msd}(c) and \ref{fig:results-two-msd}(d)], $\left \langle \mathbf{r}^{2}(t) \right \rangle$ initially indicates the normal diffusion, following Eq.~\eqref{eq:results-two-short-r2-p} (dashed line).
Then, as $t\rightarrow\infty$, it again converges to Eq.~\eqref{eq:results-two-long-r2} (dash-dotted line).
Here, strikingly, $\left \langle \mathbf{r}^{2}(t) \right \rangle$ represents the transient hyperdiffusive scaling $\beta(t)>2$.
It originates from the particles gradually transiting the dynamic states from the localization to ballistic motion via $\psi_{p}(t)$.
We manipulate the hyperdiffusive scaling by altering the PDF of localization times in Fig.~\ref{fig:results-two-hyperdiffusion}(c).
However, this hyperdiffusion is concealed at the equilibrium [see Fig.~\ref{app:fig:two-beta}(a)].
After traversing the peak, at intermediate times [Fig.~\ref{fig:results-two-msd}(c)-(ii)], the curve with $\alpha_{1}=-0.5$ (green) reduces to the subdiffusive regime, whereas the curve with $\alpha_{1}=0.9$ (magenta) attains the second summit of $\beta(t)\approx1.6$.
This complicated time-varying property of $\beta(t)$ and its dependence on the initial state are associated with site-dependent dynamics, such as HDPs \cite{cherstvy2013population}.

In Eqs.~\eqref{eq:results-two-msd-a}--\eqref{eq:results-two-msd-eq}, we can rewrite $\left \langle \mathbf{r}^{2}(t) \right \rangle/v^2$ with respect to a timescale $D/v^2$.
Accordingly, the anomalous exponent is expressed in the form of $\beta(t;D/v^2)$.
Then, in Fig.~\ref{fig:results-two-msd}(c)-(i), we find that $D/v^2$ controls the short-time behavior of $\beta(t)$ such that a smaller $D/v^2$ (dash-dotted line) strengthens the hyperdiffusive scaling $\beta(t)\approx2.8$ [see also Fig.~\ref{app:fig:two-beta}(c)].
However, the second summit of $\beta(t)\approx1.9$ in Fig.~\ref{fig:results-two-msd}(c)-(iii) arises due to $\alpha_{1}=0.99$.

Compared to the curves with $\alpha_{1}=0$ (red), $\alpha_{1}$ controls $\tilde{\tau}_{1}^{(+)}$ to accelerate or delay the recovery toward the Fickian diffusion, as shown in the solid lines in the top panels of Figs.~\ref{fig:results-two-msd}(a) and \ref{fig:results-two-msd}(c).
Additionally, $\alpha_{1}$ is responsible for $\tilde{D}_{\mathrm{eff}}$ since it determines the directional rigidity of consecutive ballistic motions.
However, $\alpha_{1}$ is irrelevant to $\beta(t)$ at short times [Figs.~\ref{fig:results-two-msd}(a)-(i) and \ref{fig:results-two-msd}(c)-(i)], which is consistent with the short-time PDF [Eq.~\eqref{eq:results-two-short-pxt}] and MSDs [Eqs.~\eqref{eq:results-two-short-r2-a} and \eqref{eq:results-two-short-r2-p}].
In other words, the physical mechanism underlying the superdiffusion in Fig.~\ref{fig:results-two-msd} differs for the short- and long-time regimes.

\begin{figure*}[!t]
\centering
\includegraphics[keepaspectratio=true, width=17.8cm, height=100cm]{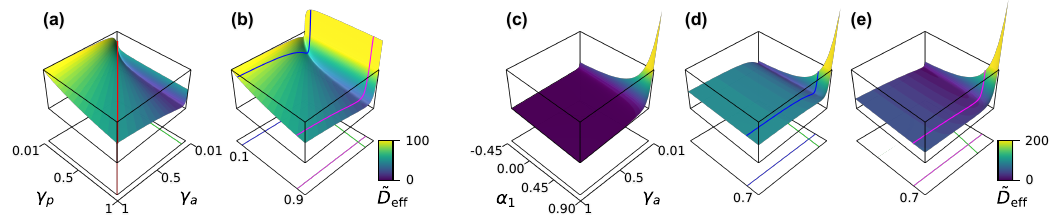}  
\caption{
Effective diffusivity $\tilde{D}_{\mathrm{eff}}$ in two parameter spaces: (a), (b) $(\gamma_{a}, \gamma_{p})$ and (c)--(e) $(\gamma_{a}, \alpha_{1})$ space for fixed $v=1$ and $D=100$; or $D=0$ in (c).
(a), (b) $\tilde{D}_{\mathrm{eff}}$ for $\alpha_{1}=0$ (a) and $0.7$ (b).
It increases as $\alpha_{1}$ approaches one.
The red line in (a) represents $\tilde{D}_{\mathrm{eff}}$ of RTPs without directional memory (see Sec.~\ref{sec:results-two-wo-memory}), satisfying $\gamma_{a}=\gamma_{p}$.
(c)--(e) $\tilde{D}_{\mathrm{eff}}$ for $\gamma_{p}=0.1$ (c), (d) and $0.9$ (e).
The local dispersion originating from the passive state with $D>0$ increases $\tilde{D}_{\mathrm{eff}}$ compared to the case of $D=0$; namely, RTPs with rests (see Sec.~\ref{sec:results-two-with-rests}).
The blue and magenta lines in (b), (d), and (e) indicate the same parameter conditions, respectively.
}
\label{fig:results-two-diffusivity}
\end{figure*}

\subsection{\label{sec:results-two-deff}Effective diffusivity}

We investigate $\tilde{D}_{\mathrm{eff}}$, especially in two parameter spaces: $(\gamma_{a}, \gamma_{p})$ and $(\gamma_{a}, \alpha_{1})$ space.
First, we plot $\tilde{D}_{\mathrm{eff}}$ on the $(\gamma_{a}, \gamma_{p})$ plane in Figs.~\ref{fig:results-two-diffusivity}(a) and \ref{fig:results-two-diffusivity}(b) for $\alpha_{1}=0$ and $0.7$, respectively.
Rewriting Eq.~\eqref{eq:results-two-diffusivity} as $\tilde{D}_{\mathrm{eff}}=\pi_{p}D+\pi_{a}^2v^2\tau_{1}/2$, we infer that the first term $\pi_{p}D$ is dominant if $\gamma_{a}\gg  \gamma_{p}$; in contrast, if $\gamma_{a}\ll \gamma_{p}$, the second term $\pi_{a}^2v^2\tau_{1}/2$ leads $\tilde{D}_{\mathrm{eff}}$, whereby we consistently interpret Figs.~\ref{fig:results-two-diffusivity}(a) and \ref{fig:results-two-diffusivity}(b).
Also, we consider a critical point $\gamma_{a}^{*}=v^2[\sqrt{2\gamma_{p}(1-\alpha_{1})D/v^2+1}+1]/[2(1-\alpha_{1})D]$, which satisfies $\partial\tilde{D}_{\mathrm{eff}}/\partial{\gamma_{a}}|_{\gamma_{a}^{*}}=0$.
Then, as $\alpha_{1}$ becomes larger in Fig.~\ref{fig:results-two-diffusivity}(b), we observe a significant increase of $\tilde{D}_{\mathrm{eff}}$ within the region $\gamma_{a}<\gamma_{a}^{*}$ (green line) compared to the same domain in Fig.~\ref{fig:results-two-diffusivity}(a).
On the contrary, we can hardly extend $\tilde{D}_{\mathrm{eff}}$ through the directional persistence of active transport ($\alpha_{1}>0$) when $\gamma_{a}>\gamma_{a}^{*}$.

For $\tilde{D}_{\mathrm{eff}}$ on the $(\gamma_{a}, \alpha_{1})$ plane [Figs.~\ref{fig:results-two-diffusivity}(c)--\ref{fig:results-two-diffusivity}(e)], we confirm that $\tilde{D}_{\mathrm{eff}}$ is positively correlated with $\alpha_{1}$, as expected in Eq.~\eqref{eq:results-two-diffusivity}.
In particular, within the region $\gamma_{a}<\gamma_{a}^{*}$ (green line), we can generate a dispersion higher than the passive fluctuation $D=100$ since the mean duration of the active state is sufficiently long to exploit the directional memory via $\alpha_{1}>0$.
Also, $\tilde{D}_{\mathrm{eff}}$ in Figs.~\ref{fig:results-two-diffusivity}(d) and \ref{fig:results-two-diffusivity}(e) manifests an elevation through the extra dispersion originating from the passive state with $D>0$ compared to Fig.~\ref{fig:results-two-diffusivity}(c), where we intermittently force particles to halt by using the condition $D=0$.

\subsection{\label{sec:results-two-wo-memory}RTP without directional memory}

As a special case, we consider the two-state RTPs operated following the uniform distribution $Q(\phi)=1/(2\pi)$.
The MSDs are easily derived by inserting $\alpha_{1}=0$ into Eqs.~\eqref{eq:results-two-msd-a}--\eqref{eq:results-two-msd-eq} under the condition $\gamma_{a} \neq \gamma_{p}$.
On the other hand, at $\gamma_{a}=\gamma_{p}$ and $\alpha_{1}=0$, we should take the limit $\epsilon_{1}\rightarrow0$ on Eqs.~\eqref{eq:results-two-msd-a} and \eqref{eq:results-two-msd-p} to attain their asymptotic behaviors, which read
\begin{align}
\left \langle \mathbf{r}^{2}(t) \right \rangle
    =(2D+\frac{v^{2}}{\gamma_{a}})t 
    +\frac{2D\gamma_{a}+v^{2}}{2\gamma_{a}^{2}}(e^{-2\gamma_{a} t}-1),
    \label{eq:results-two-msd-uniform-a}
    \\
    \left \langle \mathbf{r}^{2}(t) \right \rangle
    =(2D+\frac{v^{2}}{\gamma_{a}})t
    -\frac{2D\gamma_{a}+v^{2}}{2\gamma_{a}^{2}}(e^{-2\gamma_{a} t}-1)
    \nonumber \\
    +\frac{2v^{2}}{\gamma_{a}^{2}}(e^{-\gamma_{a} t}-1),
    \label{eq:results-two-msd-uniform-p}
\end{align}
where we set the initial state $i=a$ and $p$, respectively.
At the equilibrium, we reach
\begin{equation}
    \left \langle \mathbf{r}^{2}(t) \right \rangle
    =(2D+\frac{v^{2}}{\gamma_{a}})t 
    +\frac{v^{2}}{\gamma_{a}^{2}}(e^{-\gamma_{a} t}-1)
    \label{eq:results-two-msd-uniform-eq}
\end{equation}
by approximating Eq.~\eqref{eq:results-two-msd-eq}.
Here, we can rewrite Eq.~\eqref{eq:results-two-msd-uniform-eq} in a dimensionless unit such that $\left \langle \bar{\mathbf{r}}^{2}(\bar{t}) \right \rangle=(2\bar{D}+1)\bar{t} + e^{-\bar{t}} - 1$, introducing $\bar{t}=\gamma_{a} t$, $\bar{\mathbf{r}}=\gamma_{a}\mathbf{r}/v$, and $\bar{D}=D \gamma_{a}/v^{2}$.
Then, we can interpret the MSD through a single rescaled parameter $\bar{D}$.

\begin{figure}[!t]
\centering
\includegraphics[keepaspectratio=true, width=8.6cm, height=100cm]{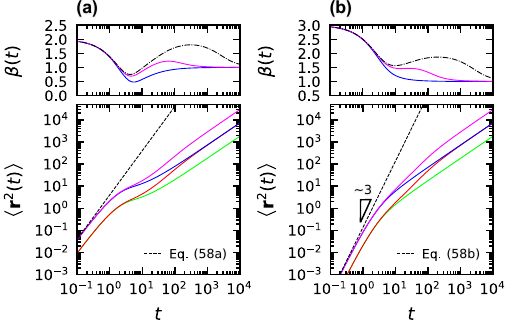}  
\caption{
Anomalous exponent $\beta(t)$ and MSDs $\left \langle \mathbf{r}^2(t) \right \rangle$ versus time $t$ when (a) the initial state $i=a$ and (b) $i=p$ for $\gamma_{a}=1$, $\gamma_{p}=0.1$, and $D=0$.
The parameters used are when $v=1$, $\alpha_{1}=0$ (green) and $0.75$ (red); when $v=2$, $\alpha_{1}=0$ (blue) and $0.75$ (magenta).
(a), (b) (top) $\beta(t)$.
The dash-dotted line indicates the case of $v=2$ and $\alpha_{1}=0.99$.
(bottom) $\left \langle \mathbf{r}^2(t) \right \rangle$.
Notably, (b) initially displays $\beta(t)=3$, but the hyperdiffusion disappears at the equilibrium [see Fig.~\ref{app:fig:two-beta}(b)].
}
\label{fig:results-two-rtp-with-rest}
\end{figure}

\subsection{\label{sec:results-two-with-rests}RTP with rests}

In Eq.~\eqref{eq:results-two-short-r2-p}, we infer that as $D\rightarrow 0$, $\left \langle \mathbf{r}^{2}(t) \right \rangle$ evolves faster than the Brownian scaling $\propto t$ at short times.
Motivated by this observation, we investigate our two-state model satisfying $D=0$ in detail.
Here, particles pause the ballistic motion intermittently; thus, we refer to this model as an RTP with rests analogously to a L\'evy walk with rests \cite{zaburdaev2015levy, klafter1994levy}.
In this case, we can directly obtain the MSDs by substituting $D=0$ into Eqs.~\eqref{eq:results-two-msd-a}--\eqref{eq:results-two-msd-eq}.

\begin{figure*}[!t]
\centering
\includegraphics[keepaspectratio=true, width=17.8cm, height=100cm]{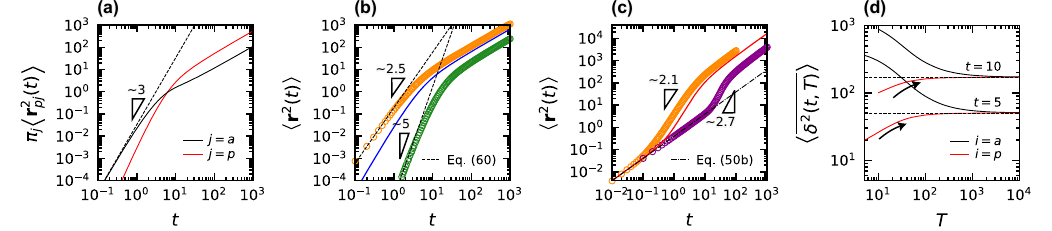}  
\caption{
Characteristics of hyperdiffusion when (a), (b) $D=0$ and (c), (d) $D>0$.
(a) MSDs $\pi_{j}\left \langle \mathbf{r}_{pj}^{2}(t) \right \rangle$ of particles initially in the passive state and observed in the state $j\in\{a,p\}$ at time $t$.
The dashed line [Eq.~\eqref{eq:results-two-rtp-with-rest-short-r2-pa}] coincides with Eq.~\eqref{eq:results-two-rtp-with-rest-short-r2-p}.
The parameters used are $\gamma_{a}=0.5$, $\gamma_{p}=0.1$, and $v=1$.
(b), (c) $\left \langle \mathbf{r}^{2}(t) \right \rangle$ when the waiting-time distribution of the passive state is the Gamma distribution $G(t;\lambda,\gamma_{p})$ for $\lambda=0.5$ (yellow), $3$ (green), and $5$ (magenta).
The solid lines in (b) and (c) are the MSDs of our Poissonian RTPs using the exponential distribution ($\lambda=1$).
The parameters used are $\gamma_{a}=0.5$, $\gamma_{p}=0.1$, and $v=1$ in (b); or $v=5$ and $D=0.1$ in (c).
(d) ETAMSDs against $T$ for fixed $\gamma_{a}=0.1$, $\gamma_{p}=0.01$, $v=5$, and $D=0.1$.
The system manifests ergodicity at the equilibrium (dashed line); however, it implies an acceleration process when the initial state $i=p$ (arrow).
We set $\alpha_{1}=0$ in (a), (b), (c), and (d) since the hyperdiffusion disappears at the long-time regime.
}
\label{fig:results-two-hyperdiffusion}
\end{figure*}


We compute the asymptotic forms of Eqs.~\eqref{eq:results-two-msd-a} and \eqref{eq:results-two-msd-p} in the limit $t \rightarrow 0$ and $D \rightarrow 0$, resulting in
\begin{subnumcases}{
    \lim_{t\rightarrow 0}\left \langle \mathbf{r}^{2}(t) \right \rangle =
}
    v^{2} t^{2}, & $i=a$, 
    \label{eq:results-two-rtp-with-rest-short-r2-a}\\
    \frac{1}{3}v^{2}\gamma_{p} t^{3}, & $i=p$,
    \label{eq:results-two-rtp-with-rest-short-r2-p}
\end{subnumcases}
respectively.
We highlight that Eq.~\eqref{eq:results-two-rtp-with-rest-short-r2-p} grows cubically with time, exhibiting the scaling $\propto t^3$ \cite{comments-pause2016} [Fig.~\ref{fig:results-two-rtp-with-rest}(b)].
On the other hand, in the limit $t \rightarrow \infty$ and $D \rightarrow 0$, we recover the Fickian diffusion in Eq.~\eqref{eq:results-two-long-r2} with $\tilde{D}_{\mathrm{eff}}=\pi_{a}v^2/[2\gamma_{a}(1-\alpha_{1})]$ for all initial states.
In Fig.~\ref{fig:results-two-rtp-with-rest}, we point out that the model carries intrinsic heterogeneity in dynamics since the short-time behavior drastically deviates depending on the initial state.
In contrast to the case of $D>0$, when $v$ is multiplied by a real-valued $c>0$, $\left \langle \mathbf{r}^{2}(t) \right \rangle$ proportionally increases with a factor of $c^2$.
Also, since $\beta(t)$ does not depend on $v$, the top panels in Fig.~\ref{fig:results-two-rtp-with-rest} display the perfect overlap for the curves with the equal $\alpha_{1}$.

To trace the origin of hyperdiffusion, we analyze $\left \langle \mathbf{r}_{p}^{2}(t) \right \rangle$ based on the final state $j$ that particles display at time $t$.
In particular, we calculate $\pi_{j}\left \langle \mathbf{r}_{pj}^{2}(t) \right \rangle=\int_{\mathbb{R}^{2}} \mathbf{r}^{2}P_{pj}(\mathbf{r},t) d^{2}\mathbf{r}$ using $\hat{P}_{pj}(\mathbf{k},s)$ in Eq.~\eqref{app:eq:two-pks-pj} for each state $j\in\{a,p\}$.
Since the scaling $\beta(t)=3$ occurs at short times, we take the limit $\alpha_{1}\rightarrow0$ (and $D\rightarrow0$ for RTPs with rests) on $\hat{P}_{pj}(\mathbf{k},s)$.
After performing the inverse Laplace transform on $\left \langle \mathbf{r}_{pj}^{2}(s) \right \rangle$, we obtain Eqs.~\eqref{app:eq:two-r2-pa} and \eqref{app:eq:two-r2-pp}.
In the limit $t\rightarrow0$, $\left \langle \mathbf{r}_{p}^{2}(t) \right \rangle$ splits into
\begin{subnumcases}{
    \lim_{t\rightarrow 0}\pi_{j}\left \langle \mathbf{r}_{pj}^{2}(t) \right \rangle =
}
    \frac{1}{3}v^{2}\gamma_{p} t^{3}, & $j=a$, 
    \label{eq:results-two-rtp-with-rest-short-r2-pa}\\
    \frac{1}{12}\gamma_a\gamma_pv^2t^4, & $j=p$.
    \label{eq:results-two-rtp-with-rest-short-r2-pp}
\end{subnumcases}
The scaling behavior of Eq.~\eqref{eq:results-two-rtp-with-rest-short-r2-pa} at short times is identical to that of Eq.~\eqref{eq:results-two-rtp-with-rest-short-r2-p} [see Fig.~\ref{fig:results-two-hyperdiffusion}(a)].
In other words, hyperdiffusion is caused by particles initially localized and gradually transiting their dynamic modes to the running states.
Even though a single particle exhibits the discontinuous transition of $\left \langle \mathbf{r}_{p}^{2}(t) \right \rangle$ (that is, zero to $v^2t^2$), the system evaluated in an ensemble-averaged manner displays the continuous increase of $\left \langle \mathbf{r}_{p}^{2}(t) \right \rangle \propto t^3$.
Thus, we can regard it as an \textit{acceleration process} driven by particles gradually escaping the localization following the exponential waiting-time distribution $\psi_{p}(t)$.

More generally, we alter the transition rate of $p\rightarrow a$ using a Gamma distribution $G(t;\lambda,\gamma_{p})=\gamma_{p}^{\lambda}/\Gamma (\lambda)t^{\lambda-1}\exp(-\gamma_{p}t)$, where $\Gamma (\lambda)$ is a Gamma function for $\lambda>0$.
Consider the first state transition of $p\rightarrow a$.
When the observation time $t$ falls in the interval $X_{1}\leq t \leq X_{1}+Y_{1}$, where $X_{1}$ and $Y_{1}$ denote the first duration of $p$ and $a$, respectively, we can write $\left \langle \mathbf{r}^{2}(t) \right \rangle=\int_{0}^{t}v^2(t-X_{1})^2G(X_{1})dX_{1}$.
In the short-time regime, we reach
\begin{equation}
    \lim_{t\rightarrow 0}\left \langle \mathbf{r}^{2}(t) \right \rangle
    =\frac{2}{\lambda(\lambda+1)(\lambda+2)\Gamma(\lambda)}v^2\gamma_{p}^{\lambda}t^{2+\lambda},
\end{equation}
where we recover the MSD of our Poissonian RTPs [Eq.~\eqref{eq:results-two-rtp-with-rest-short-r2-p}] for $\lambda=1$.
In other words, $\beta(t)=3$ at short times represents the Poissonian feature of the RTPs with rests.
In contrast, Fig.~\ref{fig:results-two-hyperdiffusion}(b) shows the numerical results (circles) of non-Poissonian RTPs for $\lambda\neq1$.
We confirm that by regulating the transition rate of $p\rightarrow a$, the dispersion speed of the system indeed changes following the scaling $\propto t^{2+\lambda}$.

Lastly, we remark on the MSD of the RTPs with rests when the model follows $Q(\phi)=1/(2\pi)$.
In the limit $\alpha_{1} \rightarrow 0$ and $D\rightarrow0$, Eq.~\eqref{eq:results-two-msd-eq} becomes
\begin{equation}
    \left \langle \mathbf{r}^{2}(t) \right \rangle
    =\frac{2v^2\pi_{a}}{\gamma_{a}}t+\frac{2v^2\pi_{a}}{\gamma_{a}^2}(e^{-\gamma_{a}t}-1)
    \label{eq:results-two-rtp-with-rest-msd-eq-uniform}
\end{equation}
for particles initially residing at the equilibrium.
Compared to the MSD of the ABPs [Eq.~\eqref{eq:abp-msd}], $\gamma_{a}$ replaces $D_{\mathrm{R}}$, and $v$ multiplied by $\sqrt{\pi_{a}}$ manifests an effective decrease of the speed.
Moreover, taking the limit $\gamma _{p}\rightarrow \infty$ on Eq.~\eqref{eq:results-two-rtp-with-rest-msd-eq-uniform}, it recovers the MSD of the conventional RTPs [Eq.~\eqref{eq:rtp-convention-msd}].

\section{\label{sec:discussion}Discussion}

We discuss the hyperdiffusion phenomenon further, especially for particles experiencing a local fluctuation $D>0$.
For a sufficiently small $D/v^2$, we have reported that the scaling $\beta(t)>2$ arises in the time domain $t\lesssim\tilde{\tau}_{1}$ selectively to particles initially in the passive state [see Fig.~\ref{app:fig:two-beta}(c)].
Considering the Gamma distribution $G(t;\lambda\neq1, \gamma_{p})$ as the waiting-time distribution of the passive state, the transition rate of $p\rightarrow a$ regulates the anomalous exponent and hyperdiffusive regime of $\left \langle \mathbf{r}^{2}(t) \right \rangle$, as depicted in Fig.~\ref{fig:results-two-hyperdiffusion}(c).
Yet, the scaling difference to the Poissonian case (red solid line) is not significant compared to the RTPs with rests in Fig.~\ref{fig:results-two-hyperdiffusion}(b).

We have argued that the hyperdiffusion is caused by the particles \textit{gradually} escaping the localized states.
In other words, if the complex system traps the particles intermittently and releases them in a growing manner as time passes, we may suspect the scaling $\beta(t)>2$.
Relevant theoretical studies have been conducted \cite{lu2007state, siegle2010markovian, siegle2010origin} where a generalized Langevin equation (GLE) with various memory kernels is applied.
For instance, Ref.~\cite{lu2007state} reported that super-Ohmic damped particles in a tilted periodic potential display $\beta(t)>2$ while they coexist in locked and running states.
Similarly, Ref.~\cite{siegle2010origin} produced the enhanced anomaly via the GLE with long-ranged velocity correlation under a tiled washboard potential.
More generally, we remark on the experimental study of Ref.~\cite{levi2012hyper}, where the hyperdiffusion of light is observed while it propagates inside an evolving disordered system.
It is the first experimental evidence of accelerated transport evoked from non-static disordered media beyond the conventional expectation of Anderson localization \cite{anderson1958absence, peccianti2012beyond}.
Hence, it would be interesting to experimentally discover this accelerated transport in biophysics.
Moreover, developing a theoretical model to shorten or enlarge the intermittent hyperdiffusive regime could be another intriguing future work.

We investigate whether the two-state model has an ergodic property.
For quantitative analysis, we numerically calculate the ensemble-averaged TAMSD (ETAMSD) $\left \langle \overline{\delta^2(t,T)} \right \rangle$ with respect to the total measurement time $T$ \cite{klafter2011first}.
In Fig.~\ref{fig:results-two-hyperdiffusion}(d), the equilibrium case (dashed lines) satisfies the ergodicity for all $T$, whereas the other two cases (black and red solid lines) manifest ergodicity breaking at short times.
In particular, for particles initially in the passive state (red solid lines), ETAMSD increases opposite to the well-known aging curve \cite{klafter2011first, zaburdaev2015levy}, which consistently implies the acceleration process.


\section{\label{sec:conclusion}Conclusion}

Using the coupled CTRW and Monte Carlo simulations, we have studied non-interacting Poissonian RTPs with(out) passive Brownian motion in the $2$D plane, where the propagation direction is manipulated by a circular distribution $Q(\phi)$.
We conclude with three crucial remarks:
\begin{enumerate}[(i)]
    \item For both one-/two-state RTPs, the displacement moment $\left \langle \mathbf{r}^{2\mu}(t) \right \rangle$ depends on the finite cosine moments $\alpha_{\nu\leq\mu}$ of $Q(\phi)$ for non-negative integers $\nu,\mu$.
    In particular, the first cosine moment $\alpha_{1}$ regulates the effective diffusivity $D_{\mathrm{eff}}$/$\tilde{D}_{\mathrm{eff}}$ and timescale $\tau_{1}$/$\tilde{\tau}_{1}$ to recover Gaussian PDF at long times.
    On the other hand, the higher moments $\alpha_{\nu\ge2}$ only affect the PDF at intermediate times.
    \item For one-state RTPs, we have suggested the angular distribution of velocity reorientation $\mathcal{Q}(\Theta,t)$ as the high-resolved stochastic quantity because it contains the whole information of $Q(\phi)$ (tumbling patterns).
    By extracting the time-dependent cosine moment $\mathcal{M}_{\mu}(t)$ from $\mathcal{Q}(\Theta,t)$, we have distinguished SPPs, such as ABPs and RTPs with various $Q(\phi)$.
    \item For two-state RTPs, we have unveiled $\left \langle \mathbf{r}^{2}(t) \right \rangle$ at all time regimes depending on initial states.
    The MSDs display an intricate time evolution, e.g., a rapid transition from sub to superdiffusion at intermediate times.
    The physical mechanism underlying superdiffusion differs at short and long times; the local ballistic motion and long-ranged directional persistence, respectively.
    Remarkably, for a sufficiently small $D/v^2$, the hyperdiffusive scaling $2<\beta(t)\leq3$ is observed at short and intermediate times for particles initially in the passive state, which even manifests transient ergodicity breaking.
\end{enumerate}

Finally, we stress that the hyperdiffusion originates from the localized particles that gradually transit the dynamic mode to the ballistic motion.
Also, the heavy dependence of MSDs on initial states implies a connection of our model to HDPs, where the geometric characteristics of media can determine a tracer’s initial condition.
Conclusively, we provide a realistic SPP model that is extendable to $n$-state cases and widely applicable for localized active processes in crowded and disordered systems.


\begin{widetext}
\appendix

\section{\label{app:sec:transforms}TRANSFORMS AND CONVOLUTIONS}

To deal with the transport equations of the CTRW, we utilize the Fourier and Laplace transforms, defined as
\begin{equation}
    \mathcal{F}\{f(\mathbf{r},t)\}
    :=\int_{\mathbb{R}^2} e^{-i\mathbf{k}\cdot\mathbf{r}} f(\mathbf{r},t) d^{2}\mathbf{r},
    \hspace{0.5cm}
    \mathcal{L}\{f(\mathbf{r},t)\}
    :=\int_{0}^{\infty} e^{-st} f(\mathbf{r},t) dt,
\end{equation}
respectively. In the combined form, we write
\begin{equation}
    \hat{f}(\mathbf{k},s)
    :=\mathcal{FL}\{f(\mathbf{r},t)\}
    :=\int_{0}^{\infty}\int_{\mathbb{R}^2} e^{-i\mathbf{k}\cdot\mathbf{r}} e^{-st} f(\mathbf{r},t) d^{2}\mathbf{r}dt,
\end{equation}
where we abbreviate the Fourier-Laplace transform of $f(\mathbf{r},t)$ as $\hat{f}(\mathbf{k},s)$.

A circular convolution for two functions $f(\mathbf{r})=f(r, \theta)$ and $g(\mathbf{r})=g(r, \theta)$ is defined as \cite{baddour2011two}
\begin{equation}
    f(\mathbf{r})*_{\theta}g(\mathbf{r})=\frac{1}{2\pi}\int_{-\pi}^{\pi}f(r,\phi)g(r,\theta-\phi)d\phi.
    \label{app:eq:circular-convolution-definition}
\end{equation}
It is especially beneficial when treating the variables $r$ and $\theta$ separately, such as Eq.~\eqref{eq:one-zeta-jacobi-anger} in the main text.

\section{\label{app:sec:circular}CIRCULAR DISTRIBUTIONS AND RELATED OBSERVABLES}

We recap the mathematical properties of two circular distributions: cardioid and wrapped Cauchy distributions.
Both distributions are characterized by two parameters, that is, a location parameter $\mu$ and a concentration parameter $\kappa$, and satisfy reflective symmetry with respect to $\mu$.
Specifically, the cardioid distribution is given as
\begin{equation}
    Q(\phi;\mu, \kappa) = \frac{1}{2\pi}
    \big[
        1+2\kappa\cos(\phi-\mu)
    \big]
    \label{app:eq:cardioid}
\end{equation}
for $\mu \in [-\pi, \pi)$ and $\left | \kappa \right | \leq 1/2$.
On the other hand, the wrapped Cauchy distribution is defined as
\begin{equation}
    Q(\phi;\mu, \kappa) = \frac{1}{2\pi} 
    \big[
        1+2\sum_{n=1}^{\infty}\kappa^{n}\cos n(\phi-\mu) 
    \big]
    = \frac{1}{2\pi}\frac{1-\kappa^{2}}{1+\kappa^{2}-2\kappa\cos(\phi-\mu)}
    \label{app:eq:wrapped-cauchy}
\end{equation}
for $\mu \in [-\pi, \pi)$ and $0 \leq \kappa<1$.
In this study, we fix $\mu=0$ for both distributions.
Last, a wrapped normal distribution reads
\begin{equation}
    Q(\phi;\mu, \sigma^{2}) = \frac{1}{2\pi} 
    \big[
        1+2\sum_{n=1}^{\infty}\exp[-\frac{\sigma^2n^2}{2}]\cos n(\phi-\mu) 
    \big]
    \label{app:eq:wrapped-normal}
\end{equation}
for $\mu \in [-\pi, \pi)$ and $\sigma>0$.
Note that $\kappa := \exp[-\sigma^2 /2]$ shortens the exponential term to $\kappa^{n^{2}}$ \cite{mardia2000directional}.

For a circular distribution $Q(\phi)$, the $\nu$th trigonometric moment $\chi_{\nu}$ is defined as
\begin{equation}
    \chi_{\nu}:= \left \langle \exp (i\nu\phi) \right \rangle=\int_{-\pi}^{\pi}e^{i\nu\phi}Q(\phi)d\phi
    \label{app:eq:chi}
\end{equation}
for an integer $\nu$.
It results in the $\nu$th cosine moment $\alpha_{\nu}$ and sine moment $\beta_{\nu}$:
\begin{equation}
    \alpha_{\nu}:= \left \langle \cos \nu\phi \right \rangle=\int_{-\pi}^{\pi}\cos\nu\phi Q(\phi) d\phi,
    \hspace{0.5cm}
    \beta{\nu}:= \left \langle \sin \nu\phi \right \rangle=\int_{-\pi}^{\pi}\sin\nu\phi Q(\phi) d\phi,
    \label{app:eq:alpha-beta}
\end{equation}
where we have a simple relation $\chi_{\nu}=\alpha_{\nu}+i\beta_{\nu}$.
Also, considering the Fourier series expansion $Q(\phi)=1/(2\pi)\sum_{n=-\infty}^{\infty}Q_{n}e^{in\phi}$, we reach $\chi_{\nu}=Q_{-\nu}$ from Eq.~\eqref{app:eq:chi}.
Moreover, if $Q(\phi)$ satisfies the reflective symmetry, we can use $Q_{\pm \nu}=\alpha_{|\nu|}$ due to $\beta_{\pm\nu}=0$.

As an angle-related observable, we compute a velocity autocorrelation function $\mathrm{VACF}(t)$, given as
\begin{equation}
    \mathrm{VACF}(t):=\left \langle \mathbf{v}(0)\cdot\mathbf{v}(t) \right \rangle,
    \label{app:eq:vacf-definition}
\end{equation}
where $\left \langle \cdots \right \rangle$ indicates the ensemble average.

\section{\label{app:sec:proof}PROOF OF EQ.~\eqref{eq:one-r2mu-qmu-relation}}

Based on the CTRW \cite{klafter2011first}, we calculate the $2\mu$th moment of displacement in the Laplace domain through
\begin{equation}
    \left \langle \mathbf{r}^{2\mu}(s) \right \rangle 
    = (-i)^{2\mu}\nabla_{\mathbf{k}}^{2\mu} \hat{P}(\mathbf{k},s)\big|_{\mathbf{k}=0}
\end{equation}
for a non-negative integer $\mu$.
However, we argue that $\hat{P}(\mathbf{k},s)$ can be replaced with the truncated form, e.g., $\hat{P}^{(\mu)}(\mathbf{k},s)$ [Eq.~\eqref{eq:one-r2mu-del}] or $\hat{P}^{(\mu)}_{i}(\mathbf{k},s)$ for each initial state $i\in\{a,p\}$ [Eq.~\eqref{eq:two-r2mu-del}] in the main text.

For the one-state case, we consider the series expansion of $f_{n}(\mathbf{k},s)$ at point $\mathbf{k}=0$.
Equation~\eqref{eq:one-fn} is explicitly written as
\begin{equation}
    f_{n}(\mathbf{k},s)
    =(i)^{-|n|} \frac{\gamma_{a}}{\sqrt{(\gamma_a+s)^2+\mathbf{k}^2v^2}}
    \big( \frac{|\mathbf{k}| v}{\sqrt{(\gamma_a+s)^2+\mathbf{k}^2v^2}+\gamma_{a}+s} \big)^{|n|} e^{in\psi},
    \label{app:eq:one-fn}
\end{equation}
where we use the property of the Bessel function of the first kind, $J_{-n}(x)=(-1)^{n}J_{n}(x)$ for an integer $n$.
Then, the lowest order term of $f_{n}$ in $\mathbf{k}$ is $|\mathbf{k}|^{|n|}$ in a straightforward way (here, we ignore coefficients).
Accordingly, using the relations
\begin{equation}
    Z_{\nu=0}^{\mu} = \sum_{n=-\mu}^{\mu} 
    f_{n}(\mathbf{k},s) Z_{n}^{\mu}(\mathbf{k},s) Q_{n} + \frac{1}{2\pi},
    \hspace{0.5cm}
    Z_{\nu\neq 0}^{\mu}=\sum_{n=-\mu}^{\mu} f_{n-\nu}(\mathbf{k},s) Z_{n}^{\mu}(\mathbf{k},s) Q_{n},
    \label{app:eq:one-matrix-equation}
\end{equation}
the lowest order term of $Z_{n}^{\mu}$ in $\mathbf{k}$ is also $|\mathbf{k}|^{|n|}$.
Thus, the first term in the expansion of $f_{n}Z_{n}^{\mu}$ becomes $|\mathbf{k}|^{|2n|}$.
Since $Z_{0}^{\mu}$ in Eq.~\eqref{app:eq:one-matrix-equation} already includes all possible terms contributing to $|\mathbf{k}|^{2\mu}$, the truncated form $\hat{P}^{(\mu)}(\mathbf{k},s)$ in Eq.~\eqref{eq:one-pks-mu} is enough for precisely calculating $\left \langle \mathbf{r}^{2\mu}(s) \right \rangle$.

Analogously, for the two-state case, we calculate the series expansion of $\tilde{f}_{n}(\mathbf{k},s)$ [Eq.~\eqref{eq:two-tilde-fn}], explicitly given as
\begin{equation}
    \tilde{f}_{n}(\mathbf{k},s)
    =(i)^{-|n|}
    \frac{\gamma_{a}\gamma_{p}}{(\sqrt{(\gamma_a+s)^2+\mathbf{k}^2v^2})(s+D\mathbf{k}^2+\gamma_{p})}
    \big( \frac{|\mathbf{k}| v}{\sqrt{(\gamma_a+s)^2+\mathbf{k}^2v^2}+\gamma_{a}+s} \big)^{|n|} e^{in\psi}.
    \label{app:eq:two-tilde-fn}
\end{equation}
Here, we again obtain $|\mathbf{k}|^{|n|}$ as the lowest order term of $\tilde{f}_{n}$ in $\mathbf{k}$.
From the relations
\begin{equation}
    \tilde{Z}_{\nu=0}^{\mu} 
    = \sum_{n=-\mu}^{\mu} 
    \tilde{f}_{n}(\mathbf{k},s) \tilde{Z}_{n}^{\mu}(\mathbf{k},s) Q_{n} 
    + \frac{1}{2\pi},
    \hspace{0.5cm}
    \tilde{Z}_{\nu\neq 0}^{\mu}
    = \sum_{n=-\mu}^{\mu} 
    \tilde{f}_{n-\nu}(\mathbf{k},s) \tilde{Z}_{n}^{\mu}(\mathbf{k},s) Q_{n},
    \label{app:eq:two-matrix-equation}
\end{equation}
we have $|\mathbf{k}|^{|n|}$ as the lowest order term of $\tilde{Z}_{n}^{\mu}$ in $\mathbf{k}$.
Thus, the first term in the expansion of $\tilde{f}_{n}\tilde{Z}_{n}^{\mu}$ becomes $|\mathbf{k}|^{|2n|}$, and $\tilde{Z}_{0}^{\mu}$ in Eq.~\eqref{app:eq:two-matrix-equation} already sums over all terms contributing to $|\mathbf{k}|^{2\mu}$ without the higher order terms for indices $|n|\geq\mu+1$.
Conclusively, $\left \langle \mathbf{r}^{2\mu}(s) \right \rangle$ can be exactly calculated using the truncated form $\hat{P}^{(\mu)}_{i}(\mathbf{k},s)$ in Eqs.~\eqref{eq:two-final-pks-a} and \eqref{eq:two-final-pks-p}.

\section{\label{app:sec:derivation}DERIVATION OF ANGULAR DISTRIBUTION $\mathbf{\mathcal{Q}(\Theta , t)}$}

We derive the angular distribution of velocity reorientation $\mathcal{Q}(\Theta , t)$ [see the definition of $\Theta$ in Eq.~\eqref{eq:vel-orientation-definition}].
Consider the RTPs traveling straight along the positive $x$\nobreakdash-axis.
We determine the following governing equations:
\begin{gather}
    \xi (\Theta , t)
    =\int_{0}^{t} dt_{1} \int_{-\pi}^{\pi} \xi(\Theta  - \phi, t-t_{1})Q(\phi) \psi_{a}(t_{1})d\phi 
    +\delta(\Theta )\psi_{a}(t),
    \label{app:eq:one-vel-orientation-zeta} \\
    \mathcal{Q}(\Theta , t)
    =\int_{0}^{t} dt_{1}\int_{-\pi}^{\pi} \xi(\Theta  - \phi, t-t_{1})Q(\phi) \Psi_{a}(t_{1})d\phi 
    +\delta(\Theta )\Psi_{a}(t),
    \label{app:eq:one-vel-orientation-pdf}
\end{gather}
where $\xi (\Theta , t)$ indicates the PDF of finding the particles that display a directional change $\Theta(t)$ precisely at time $t$.
Here, we designate $\delta(\Theta )\psi_{a}(t)$ as an initial condition.
Performing the Laplace transform on Eqs.~\eqref{app:eq:one-vel-orientation-zeta} and \eqref{app:eq:one-vel-orientation-pdf}, we yield
\begin{equation}
    \mathcal{L}\{\xi(\Theta , t)\}
    =\sum_{n=-\infty}^{\infty} \big[\xi_{n}Q_{n}\psi_{a}(s)+\frac{1}{2\pi}\psi_{a}(s) \big] e^{in\Theta },
    \hspace{0.5cm}
    \mathcal{L}\{\mathcal{Q}(\Theta , t)\}
    =\sum_{n=-\infty}^{\infty} \big[\xi_{n}Q_{n}\Psi_{a}(s)+\frac{1}{2\pi}\Psi_{a}(s) \big] e^{in\Theta },
\end{equation}
respectively, where we exploit the circular convolution [Eq.~\eqref{app:eq:circular-convolution-definition}] and Fourier series expansions $Q(\phi)=1/(2\pi)\sum_{n=-\infty}^{\infty} Q_{n}e^{in\phi}$ and $\delta(\Theta )=1/(2\pi)\sum_{n=-\infty}^{\infty}e^{in\Theta }$.
Finally, we reach
\begin{equation}
    \mathcal{L}\{\mathcal{Q}(\Theta , t)\}
    =\frac{1}{2\pi}\sum_{n=-\infty}^{\infty}\frac{\Psi_{a}(s)}{1-\psi_{a}(s)Q_{n}}e^{in\Theta }.
\end{equation}
Additionally, inserting the exponential distribution $\psi_{a}(t)$ and symmetric circular distribution $Q(\phi)$, we find
\begin{equation}
    \mathcal{L}\{\mathcal{Q}(\Theta , t)\}
    =\frac{1}{2\pi} \big [\frac{1}{s}+2\sum_{n=1}^{\infty}\frac{1}{s+\gamma_{a}-\gamma_{a}\alpha_{n}} \cos n\Theta \big ].
    \label{app:eq:one-vel-orientation-laplace}
\end{equation}
After performing the inverse Laplace transform on Eq.~\eqref{app:eq:one-vel-orientation-laplace}, we can get $\mathcal{Q}(\Theta , t)$ in the main text [Eq.~\eqref{eq:results-one-vel-orientation-pdf}].

\section{\label{app:sec:functions-one}SUPPLEMENTARY FUNCTIONS FOR ONE-STATE RTPS}

We address the propagator of the conventional RTPs whose transport equations are given in Eq.~\eqref{eq:rtp-convention-prt}.
First, we move the Van Hove function $P(\mathbf{r},t)$ to the Fourier-Laplace domain and yield
\begin{equation}
    \hat{P}(\mathbf{k},s)=\frac{1}{\sqrt{(\gamma_{a}+s)^2+\mathbf{k}^2v^2}-\gamma_{a}}.
    \label{app:eq:convention-pks}
\end{equation}
We recover Eq.~\eqref{app:eq:convention-pks} several times in the main text [see Eqs.~\eqref{eq:one-pks-mu} and \eqref{eq:results-one-cd-pks}].
More importantly, the Fourier-Laplace inversion $P(\mathbf{r},t)$ is already solved in Refs.~\cite{martens2012probability, santra2020run}.
In particular, through the marginalization $P(x,t)=\int_{-\infty}^{\infty} P(\mathbf{r},t)dy$, Ref.~\cite{santra2020run} obtains
\begin{equation}
    P(x,t)=e^{-\gamma_{a}t} \left[ 
            \frac{\gamma_{a}}{2v} \left(
                L_{0}(\frac{\gamma_{a}}{v}\sqrt{v^2t^2-x^2})
                + I_{0}(\frac{\gamma_{a}}{v}\sqrt{v^2t^2-x^2}) 
            \right ) + \frac{1}{\pi \sqrt{v^2t^2-x^2}}
        \right ],
    \label{app:eq:convention-pxt}
\end{equation}
where $I_{0}$ is the modified Bessel function of the first kind and $L_{0}$ is the modified Struve function.

The fourth displacement moment $\left \langle \mathbf{r}^{4}(t) \right \rangle$ is explicitly given as
\begin{align}
    \left \langle \mathbf{r}^{4}(t) \right \rangle 
    = \frac{6v^{4}}{\gamma_{a}^{4}}
    \left [
        \frac{\gamma_{a}^{2}t^{2}}{(1-\alpha_{1})^{2}}
        + \frac{e^{-\gamma_{a}(1-\alpha_{2})t}}{(\alpha_{1}-\alpha_{2})^{2}(1-\alpha_{2})^{2}} 
        - \frac{
            \alpha_{1}^{2}+2(\alpha_{2}-2)\alpha_{1}-6\alpha_{2}^2+10\alpha_{2}-3
        }{
            (1-\alpha_{1})^{4}(1-\alpha_{2})^{2}
        } \right. \hspace{3.5cm}
        \nonumber \\
        \left.
        -\frac{
            \gamma_{a}(3\alpha_{1}-2\alpha_{2}-1)te^{-\gamma_{a}(1-\alpha_{1})t}
        }{
            (1-\alpha_{1})^{3}(\alpha_{1}-\alpha_{2})
        } 
        -\frac{\gamma_{a}(\alpha_{1}-4\alpha_{2}+3)t}{(1-\alpha_{1})^{3}(1-\alpha_{2})} 
        -\frac{(9\alpha_{1}^{2}-2(7\alpha_{2}+2)\alpha_{1}+6\alpha_{2}^{2}+2\alpha_{2}+1)e^{-\gamma_{a}(1-\alpha_{1})t}}{(1-\alpha)^{4}(\alpha_{1}-\alpha_{2})^{2}}
    \right ],
    \label{app:eq:one-r4}
\end{align}
which has already been addressed in Refs.~\cite{villa2020run} and \cite{sevilla2020two}.
Additionally, we here remark on the non-Gaussianity
\begin{align}
    \mathrm{NG}(t) = \frac{1}{
        (\gamma_{a}(1-\alpha_{1})t+e^{-\gamma_{a}(1-\alpha_{1})t}-1)^2
    }
    \times
    \left [
        -e^{-2\gamma_{a}(1-\alpha_{1})t} 
        -\frac{\gamma_{a}(1-\alpha_{1})}{(\alpha_{1}-\alpha_{2})}
            (5\alpha_{1}-4\alpha_{2}-1)te^{-\gamma_{a}(1-\alpha_{1})t}
    \right. \hspace{1.25cm} 
    \nonumber \\
        -\frac{1}{(\alpha_{1}-\alpha_{2})^2}
            (7\alpha_{1}^2-2(5\alpha_{2}+2)\alpha_{1}+4\alpha_{2}^2+2\alpha_{2}+1) e^{-\gamma_{a}(1-\alpha_{1})t}
        +\frac{(1-\alpha_{1})^4}{(\alpha_{1}-\alpha_{2})^2(1-\alpha_{2})^2} e^{-\gamma_{a}(1-\alpha_{2})t}
    \nonumber \\ \left.
        +\frac{\gamma_{a}(1-\alpha_{1})}{(1-\alpha_{2})}
            (-\alpha_{1}+2\alpha_{2}-1)t
        -\frac{1}{(1-\alpha_{2})^2}
            (\alpha_{1}^2+2(\alpha_{2}-2)\alpha_{1}-5\alpha_{2}^2+8\alpha_{2}-2)
    \right ].
    \label{app:eq:one-ng}
\end{align}
Equations~\eqref{app:eq:one-r4} and \eqref{app:eq:one-ng} agree perfectly with the numerical results of Monte Carlo simulations in Figs.~\ref{app:fig:simulations}(b) and \ref{app:fig:simulations}(c).

\section{\label{app:sec:functions-two}SUPPLEMENTARY FUNCTIONS FOR TWO-STATE RTPS}

We implement the Fourier-Laplace transform on $P_{ij}(\mathbf{r},t)$ for all possible states $i,j \in \{a, p\}$.
From Eqs.~\eqref{eq:two-prt-ii} and \eqref{eq:two-prt-ij}, we derive
\begin{gather}
    \hat{P}_{aa}(\mathbf{k},s)
    =\frac{2\pi(s+D\mathbf{k}^2+\gamma_{p})}{\gamma_{a}\gamma_{p}}\sum_{n=-\infty}^{\infty} \tilde{f}_{n}(\mathbf{k},s)\hat{\tilde{\zeta}}_{n} (\mathbf{k}, s) Q_{n},
    \hspace{0.5cm}
    \hat{P}_{ap}(\mathbf{k},s)
    =\frac{2\pi}{\gamma_{p}}\sum_{n=-\infty}^{\infty} \tilde{f}_{n}(\mathbf{k},s)\hat{\tilde{\zeta}}_{n} (\mathbf{k}, s) Q_{n},
    \label{app:eq:two-pks-aj}
    \\
    \hat{P}_{pa}(\mathbf{k},s)
    =\frac{2\pi}{\gamma_{a}}\sum_{n=-\infty}^{\infty} \tilde{f}_{n}(\mathbf{k},s)\hat{\tilde{\zeta}}_{n} (\mathbf{k}, s) Q_{n},
    \hspace{0.5cm}
    \hat{P}_{pp}(\mathbf{k},s)
    =\frac{2\pi}{s+D\mathbf{k}^2+\gamma_{p}}\hat{\tilde{\zeta}}_{0} (\mathbf{k}, s) Q_{0}.
    \label{app:eq:two-pks-pj}
\end{gather}
Using Eqs.~\eqref{app:eq:two-pks-aj} and \eqref{app:eq:two-pks-pj}, we calculate $\hat{P}_{i}(\mathbf{k},s)=\sum_{j\in\{a,p\}}\hat{P}_{ij}(\mathbf{k},s)$ and $\hat{P}(\mathbf{k},s)=\sum_{i\in\{a,p\}}\pi_{i}\hat{P}_{i}(\mathbf{k},s)$.

Let us assume $Q(\phi)$ is the cardioid distribution whose cosine moments $\alpha_{\mu\geq2}$ vanish.
Then, we can obtain $\hat{P}(\mathbf{k},s)$ in a closed form.
From Eqs.~\eqref{eq:two-final-pks-a} and \eqref{eq:two-final-pks-p}, we reach
\begin{align}
    \hat{P}_{a}(\mathbf{k},s)=
    \frac{
        -\gamma_{a}\gamma_{p}\alpha_{1}v^2\mathbf{k}^2
        \big(\gamma_{a}+h\big)
        -\big(g+\gamma_{a}+s\big)^{2}
        \big(\gamma_{a}\gamma_{p}\alpha_{1}-gh\big)
        \big(\gamma_{a}+h\big)
    }{
        \gamma_{a}\gamma_{p}\alpha_{1}v^2\mathbf{k}^2
        \big(\gamma_{a}\gamma_{p}+gh\big)
        +\big(g+\gamma_{a}+s\big)^{2}
        \big(\gamma_{a}\gamma_{p}\alpha_{1}-gh\big)
        \big(\gamma_{a}\gamma_{p}-gh\big)
    },
    \label{app:eq:two-cd-pks-a}
    \\
    \hat{P}_{p}(\mathbf{k},s)=
    \frac{
        -\gamma_{a}\gamma_{p}\alpha_{1}v^2\mathbf{k}^2
        \big(\gamma_{p}-g\big)
        -\big(g+\gamma_{a}+s\big)^2
        \big(\gamma_{a}\gamma_{p}\alpha_{1}-gh\big)
        \big(\gamma_{p}+g\big)
    }{
        \gamma_{a}\gamma_{p}\alpha_{1}v^2\mathbf{k}^2
        \big(\gamma_{a}\gamma_{p}+gh\big)
        +\big(g+\gamma_{a}+s\big)^{2}
        \big(\gamma_{a}\gamma_{p}\alpha_{1}-gh\big)
        \big(\gamma_{a}\gamma_{p}-gh\big)
    },
    \label{app:eq:two-cd-pks-p}
\end{align}
where we denote $g(\mathbf{k},s)=\sqrt{(\gamma_{a}+s)^2+\mathbf{k}^2v^2}$ and $h(\mathbf{k},s)=s+D\mathbf{k}^2+\gamma_{p}$ for $|\alpha_{1}|\leq 1/2$.
At the small $(\mathbf{k}, s)$ limit, both functions approach
\begin{equation}
    \hat{P}(\mathbf{k},s) \approx
    \frac{
        2\gamma_{a}\big(\gamma_{a}+\gamma_{p}\big)\big(1-\alpha_{1}\big)
    }{
        2D\gamma_{a}^2\big(1-\alpha_{1}\big)\mathbf{k}^2
        +\gamma_{p}v^2\mathbf{k}^2
        +2\gamma_{a}\big(\gamma_{a}+\gamma_{p}\big)\big(1-\alpha_{1}\big)s
    }.
    \label{app:eq:two-long-pks}
\end{equation}
The Fourier-Laplace inversion of Eq.~\eqref{app:eq:two-long-pks} indicates the Gaussian PDF [Eq.\eqref{eq:results-two-long-pxt}].

Using Eq.~\eqref{app:eq:two-pks-pj}, we calculate $\left \langle \mathbf{r}_{pj}^{2}(t) \right \rangle$, i.e., the MSD of particles initially in the passive state and observed in the state $j\in\{a,p\}$ at time $t$.
For each final state, $\left \langle \mathbf{r}_{pj}^{2}(t) \right \rangle=1/\pi_{j}\int_{\mathbb{R}^{2}} \mathbf{r}^{2}P_{pj}(\mathbf{r},t) d^{2}\mathbf{r}$ is satisfied, and hence
\begin{gather}
    \left \langle \mathbf{r}_{pa}^{2}(t) \right \rangle
    =\frac{2v^2}{\pi_{a}\gamma_{a}^2}
    \left [
        \frac{\gamma_a\gamma_p^2t}{(\gamma_a+\gamma_p)^2}
        +\frac{\gamma_p(\gamma_a^2-2\gamma_a\gamma_p-\gamma_p^2)}{(\gamma_a+\gamma_p)^3}
        +\frac{\gamma_a^3 te^{-t(\gamma_a+\gamma_p)}}{(\gamma_a+\gamma_p)^2}
        +\frac{(\gamma_p-\gamma_a)e^{-\gamma_a t}}{\gamma_p}
        +\frac{\gamma_a^2(\gamma_a^2+2\gamma_a\gamma_p-\gamma_p^2)e^{-(\gamma_a+\gamma_p)t}}{\gamma_p(\gamma_a+\gamma_p)^3}
    \right ],
    \label{app:eq:two-r2-pa}
    \\
    \left \langle \mathbf{r}_{pp}^{2}(t) \right \rangle
    =\frac{2v^2}{\pi_{p}\gamma_{a}^2}
    \left [
        \frac{\gamma_a^2\gamma_p t}{(\gamma_a+\gamma_p)^2}
        -\frac{\gamma_a\gamma_p(3\gamma_a+\gamma_p)}{(\gamma_a+\gamma_p)^3}
        -\frac{\gamma_a^3 te^{-t(\gamma_a+\gamma_p)}}{(\gamma_a+\gamma_p)^2}
        +\frac{\gamma_a e^{-\gamma_a t}}{\gamma_p}
        -\frac{\gamma_a^3(\gamma_a+3\gamma_p)e^{-(\gamma_a+\gamma_p)t}}{\gamma_p(\gamma_a+\gamma_p)^3}
    \right ]
    \label{app:eq:two-r2-pp}
\end{gather}
are obtained.

We stress that the dynamics of our two-state RTPs is sensitive to the initial conditions.
For example, when $\beta(t)$ is calculated for $\left \langle \mathbf{r}^{2}(t) \right \rangle$ at the equilibrium, the hyperdiffusive scaling $2<\beta(t)\leq3$ disappears, as shown in Figs.~\ref{app:fig:two-beta}(a) and \ref{app:fig:two-beta}(b).
More generally, when we vary the population of particles initially in the active state (denoted by $\varpi$), the scaling $2<\beta(t)\leq3$ is intensified as $\varpi\rightarrow0$ in Fig.~\ref{app:fig:two-beta}(c).
In other words, the hyperdiffusion cannot be detected without consideration of the initial states, even for a sufficiently small $D/v^2$ that causes the phenomenon at $\varpi=0$.

\begin{figure*}[!t]
\centering
\includegraphics[keepaspectratio=true, width=17.8cm, height=100cm]{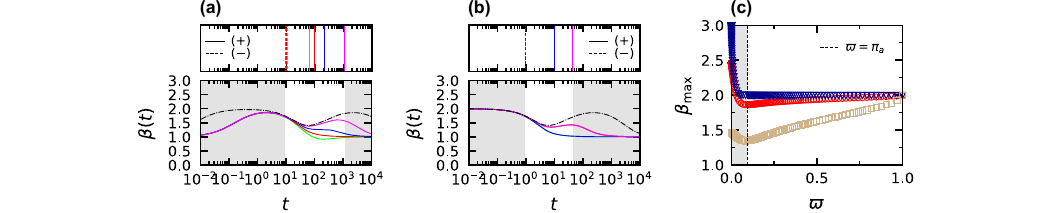}  
\caption{
Dependence of localized RTPs on initial conditions.
(a), (b) (top) Timescales $\tilde{\tau}_{(+)}$ (solid line) and $\tilde{\tau}_{(-)}$ (dashed line).
(bottom) $\beta(t)$ computed for $\left \langle \mathbf{r}^{2}(t) \right \rangle$ at the equilibrium [Eq.~\eqref{eq:results-two-msd-eq}].
Note that hyperdiffusion with the scaling $2<\beta(t)\leq3$ is hidden.
The parameters used are identical to those of Figs.~\ref{fig:results-two-msd} and \ref{fig:results-two-rtp-with-rest} for (a) and (b), respectively.
The shaded regions (gray) are divided based on $\alpha_{1}=0.9$ (magenta).
(c) $\beta_{max}:=\max_{t\lesssim\tilde{\tau}_{1}}(\beta(t))$ versus weight $\varpi$.
Here, $\beta(t)$ is computed for $\left \langle \mathbf{r}^{2}(t) \right \rangle=\varpi\times\left \langle \mathbf{r}_{a}^{2}(t) \right \rangle + (1-\varpi)\times \left \langle \mathbf{r}_{p}^{2}(t) \right \rangle$, where $\varpi$ indicates the population of particles initially in the active state.
As $\varpi\rightarrow\pi_{a}$ (dashed line), $\left \langle \mathbf{r}^{2}(t) \right \rangle$ approaches equilibrium [Eq.~\eqref{eq:results-two-msd-eq}].
The parameters used are $D=0$ (blue), $0.1$ (red), and $5$ (yellow) for fixed $\gamma_{a}=0.1$, $\gamma_{p}=0.01$, $v=5$, and $\alpha_{1}=0$.
}
\label{app:fig:two-beta}
\end{figure*}

\begin{table}[b]
\centering
\caption{Parameters in Monte Carlo simulations.}
\vspace{0.1cm}
\label{app:table:simulations}
\begin{ruledtabular}
    \begin{tabular}{cccc}
        \addlinespace[0.2ex]
        Model & Observable & Parameters & Eq(s). \\
        \addlinespace[0.2ex]
        \hline
        \addlinespace[0.2ex]
        One-state & $\hat{P}(\mathbf{k},s)$ & $\gamma_{a}=0.1, v=1$ & \eqref{eq:results-one-cd-pks} \\
        One-state & $\left \langle \mathbf{r}^{4}(t) \right \rangle$ & $\gamma_{a}=0.01, v=1$ & \eqref{app:eq:one-r4} \\
        One-state & $\mathrm{NG}(t)$ & $\gamma_{a}=0.01, v=1$ & \eqref{app:eq:one-ng} \\
        One-state & $\mathcal{M}_{\mu}(t), \mathrm{VACF}(t)$ & $\gamma_{a}=0.01, v=1$ & \eqref{eq:results-one-cosine-cd}--\eqref{eq:results-one-cosine-wn} \\
        Two-state & $\left \langle \mathbf{r}^{2}(t) \right \rangle$ & $\gamma_{a}=0.1, \gamma_{p}=0.01, v=5, D=0.1, i=a$ & \eqref{eq:results-two-msd-a} \\
        Two-state & $\left \langle \mathbf{r}^{2}(t) \right \rangle$ & $\gamma_{a}=0.1, \gamma_{p}=0.01, v=5, D=0.1, i=p$ & \eqref{eq:results-two-msd-p} \\
        \addlinespace[0.2ex]
    \end{tabular}
\end{ruledtabular}
\end{table}

\section{\label{app:sec:simulations}NUMERICAL SIMULATIONS}

We verified the equations listed in Table~\ref{app:table:simulations} with numerical simulations based on Monte Carlo methods.
Specifically, we generated $N$ trajectories $\{\mathbf{r}_{k}: k=0, 1, \dots\}$ by the following equation:
\begin{equation}
    \mathbf{r}_{k+1}=\mathbf{r}_{k}+\mathbf{1}_{A}(i_k)\cdot v\mathbf{\hat{n}}(\theta_{j})\Delta t+(1-\mathbf{1}_{A}(i_k))\cdot\sqrt{2D\Delta t}(\eta _{x}\mathbf{\hat{x}}+\eta _{y}\mathbf{\hat{y}}),
    \label{app:eq:simulations-r}
\end{equation}
where $\mathbf{1}_{A}(i_k)$ is an indicator function such that $\mathbf{1}_{A}(i_k)=1$ if state $i_k=a$, and $\mathbf{1}_{A}(i_k)=0$ if state $i_k=p$ at time $t_{k}=k\Delta t$.
For the one-state case, since $\mathbf{1}_{A}(i_k)=1$ for all $t_{k}$, the third term on the right-hand side is always ignored.
For the second term, we find $\mathbf{\hat{n}}(\theta_{j})=\cos\theta_{j}\mathbf{\hat{x}}+\sin\theta_{j}\mathbf{\hat{y}}$, i.e., an orientation of the $j$th ballistic mode, which independently evolves via
\begin{equation}
    \theta_{j+1}=\theta_{j}+\eta_{\phi},
    \label{app:eq:simulations-theta}
\end{equation}
where $\eta_{\phi}$ is drawn from a circular distribution $Q(\phi;\kappa)$ such as Eqs.~\eqref{app:eq:cardioid} and \eqref{app:eq:wrapped-cauchy}.
Initially, we make $\theta_{0}$ sampled from the uniform distribution $U(\theta)=1/(2\pi)$.
Also, we set $\theta_{j}$ to be last for a duration $X_{j}$ sampled from $\psi_{a}(t)=\gamma_{a}\exp(-\gamma_{a}t)$.
Then, introducing $S_{n}:=\sum_{j=1}^{n}X_{j}$ for an integer $n\geq1$, we can precisely write $\theta_{j}=\theta_{j(k)}$ whose $j(k)=\sup\{n: S_{n}\leq k\Delta t\}$ to update $\theta_{j(k)}$ as $t_{k}$ passes.
On the other hand, for the two-state case, due to an additional duration $Y_{j}$ sampled from $\psi_{p}(t)=\gamma_{p}\exp(-\gamma_{p}t)$, we designate $j(k)=\sup\{n: \tilde{S}_{n}\leq k\Delta t\}$, where we define $\tilde{S}_{n}:=\sum_{j=1}^{n}(X_{j}+Y_{j})$.
Additionally, white noises $\eta_{x}$ and $\eta_{y}$ in Eq.~\eqref{app:eq:simulations-r} are drawn from the standard normal distribution with zero mean and unit variance.
Last, we set initial conditions $\mathbf{r}_{0}=(0,0)$ and $S_{0}=\tilde{S}_{0}=0$; and fix the time interval to be $\Delta t=1$.

We checked that various stochastic quantities in Table~\ref{app:table:simulations} display perfect agreement with the theoretical predictions, as shown in Fig.~\ref{app:fig:simulations}.
In detail, we plotted the graphs with $N=2\times10^{4}$ realizations in Fig.~\ref{app:fig:simulations}(a), $N=5\times10^{4}$ realizations in Figs.~\ref{app:fig:simulations}(b)--\ref{app:fig:simulations}(f), and $N=10^{4}$ realizations in Figs.~\ref{app:fig:simulations}(g) and \ref{app:fig:simulations}(h).
In particular, we conducted simulations for ABPs to compare with RTPs in Fig.~\ref{app:fig:simulations}(d), which follows
\begin{equation}
    \mathbf{r}_{j+1}=\mathbf{r}_{j}+v\mathbf{\hat{n}}(\theta_{j})\Delta t,
    \hspace{0.5cm}
    \theta_{j+1}=\theta_{j}+\sqrt{2D_{\mathrm{R}}\Delta t}\eta_{\phi},
\end{equation}
where $\eta_{\phi}$ is drawn from the Gaussian distribution with zero mean and unit variance.
For the simulation source codes, refer to our \texttt{GitHub} page (\url{https://github.com/jung235}).

\begin{figure*}[!t]
\centering
\includegraphics[keepaspectratio=true, width=17.8cm, height=100cm]{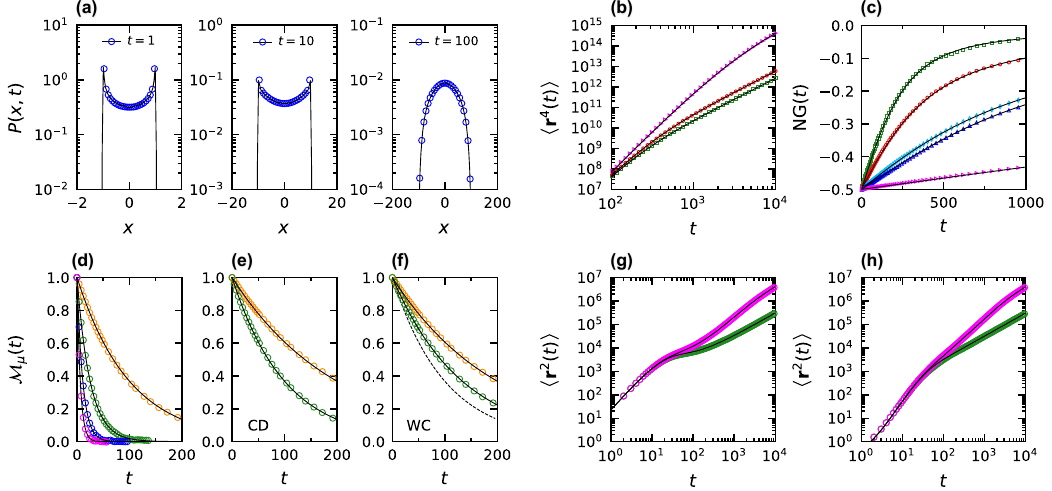}  
\caption{
Stochastic observables listed in Table~\ref{app:table:simulations}.
The simulation results (markers) and the theoretical predictions (lines) are perfectly overlaid.
(a) $P(x,t)$ when $\alpha_{1}=0.5$.
The black solid line is obtained by numerically performing the Fourier-Laplace inversion on $\hat{P}(\mathbf{k},s)$ [Eq.~\eqref{eq:results-one-cd-pks}].
(b) $\left \langle \mathbf{r}^{4}(t) \right \rangle$ for $\alpha_{1}=-0.5$ (green), $0$ (red), and $0.9$ (magenta).
(c) $\mathrm{NG}(t)$ for $\alpha_{1}=-0.5$ (green), $0$ (red), $0.5$ from the CD (blue), $0.5$ from the WC (skyblue), and $0.9$ (magenta).
(d)--(f) $\mathcal{M}_{\mu}(t)$ when the order $\mu=1$ (yellow), $2$ (green), $3$ (blue), and $4$ (magenta).
(d) ABPs for $D_{\mathrm{R}}=0.01$ and $v=1$.
(e), (f) RTPs with different $Q(\phi;\kappa)$ for $\kappa=1/2$.
The black dashed line in (f) represents $\mathcal{M}_{\infty}(t)$.
(g), (h) $\left \langle \mathbf{r}^{2}(t) \right \rangle$ for $\alpha_{1}=-0.5$ (green) and $0.9$ (magenta) when we assign the initial state $i=a$ (g) or $i=p$ (h).
}
\label{app:fig:simulations}
\end{figure*}

\end{widetext}

\nocite{*}

\bibliography{apssamp}

\end{document}